\documentclass[useAMS,usenatbib,referee]{mnras}

\usepackage{dcolumn}
\usepackage{graphicx}
\usepackage{url}
\usepackage{color}
\usepackage{pdflscape}
\usepackage{booktabs}

\usepackage[varg]{txfonts}

\graphicspath{ {eps-3/} }

\newcommand{\cm}{cm$^{-1}$}

\newcommand{\ai}{\textit{ab initio}}
\newcommand{\Ai}{\textit{Ab initio}}

\newcommand{\eqref}[1]{(\ref{#1})}

\newcommand{\2}{$_2$}
\newcommand{\X}{$X\,{}^{1}\Sigma_{g}^{+}$}
\newcommand{\A}{$A\,{}^{1}\Pi_{u}$}
\newcommand{\B}{$B\,{}^{1}\Delta_{g}$}
\newcommand{\Bp}{$B^\prime\,{}^{1}\Sigma_{g}^{+}$}
\newcommand{\Cpstate}{$C'\,{}^{1}\Pi_{g}$}
\newcommand{\Cstate}{$C\,{}^{1}\Pi_{g}$}
\newcommand{\Dstate}{$D\,{}^{1}\Sigma_{u}^{+}$}
\newcommand{\Estate}{$E\,{}^{1}\Sigma_{g}^{+}$}

\newcommand{\astate}{$a\,{}^{3}\Pi_{u}$}
\newcommand{\bstate}{$b\,^{3}\Sigma_{g}^{-}$}
\newcommand{\cstate}{$c\,^{3}\Sigma_{u}^{+}$}
\newcommand{\dstate}{$d\,{}^{3}\Pi_{g}$}
\newcommand{\estate}{$e\,{}^{3}\Pi_{g}$}

\newcommand{\ket}[1]{\vert #1 \rangle  }

\newcommand{\rme}{{\rm e}}

\newcommand{\Cv}[1]{${\mathcal C}_{#1{\rm v}}$}
\newcommand{\Dh}[1]{${\mathcal D}_{#1{\rm h}}$}

\newcommand{\allstates}{\X, \A, \B, \Bp, \astate, \bstate, \cstate\  and \dstate}

\newcommand{\duo}{{\sc Duo}}
\newcommand{\Duo}{{\sc Duo}}

\title[ExoMol XXXI: Spectroscopy of  C$_2$]{ExoMol line lists XXXI: Spectroscopy of lowest eights electronic states of C$_2$ }

\date{\today}
\author[Yurchenko et al]{\large Sergei N. Yurchenko$^1$, Istv{\'a}n Szab{\'o}$^2$, Elizaveta Pyatenko$^1$, Jonathan Tennyson$^1$   \\
$^1$Department of Physics and Astronomy, University College London, London WC1E 6BT, UK\\
$^2$Department of Chemistry, King's College London, London, SE1 1DB, UK}
\date{Accepted XXXX. Received XXXX; in original form XXXX}

\pagerange{\pageref{firstpage}--\pageref{lastpage}} \pubyear{2012}

\begin{document}


\maketitle

\begin{abstract}


Accurate line lists for the carbon dimer, C$_2$, are presented. These line lists
cover rovibronic transitions between the eight lowest electronic
states: \X, \astate, \A, \bstate, \cstate, \dstate, \B\ and \Bp.
Potential energy curves (PECs) and transition dipole moment curves are
computed on a large grid of geometries using the aug-cc-pwCVQZ-DK/MRCI
level of theory including core and core-valence correlations and
scalar relativistic energy corrections. The same level of theory is
used to compute spin-orbit and electronic angular momentum couplings.
The PECs and couplings are refined by fitting to the empirical
(MARVEL) energies of $^{12}$C$_2$ using the nuclear-motion program \Duo.
The transition dipole moment curves are represented as analytical
functions to reduce the numerical noise when computing transition line
strengths. Partition functions, full line lists, Land\'{e}-factors and
lifetimes for three main isotopologues of C$_2$
($^{12}$C$_2$,$^{13}$C$_2$ and $^{12}$C$^{13}$C) are made available in
electronic form from the CDS
(\href{www.exomol.com}{http://cdsarc.u-strasbg.fr}) and ExoMol
(\href{www.exomol.com}{www.exomol.com}) databases.

\end{abstract}

\label{firstpage}

\section{Introduction}

The C$_2$ molecule is a prominent species in a wide variety of
astrophysical sources, including comets \citep{12RoJeMa}, interstellar
clouds \citep{12HuShFe}, translucent clouds \citep{07SoWeThYo},
proto planetary nebulae \citep{10WeRoLi}, cool
carbon stars \citep{90Goorvitc}, high-temperature stars \citep{70Vartya} and the Sun
\citep{78Lambert,82BrDeGr}.  Indeed C$_2$ spectra are commonly used to
determine the $^{12}$C/$^{13}$C isotopic ratio  in carbon stars
\citep{09ZaAbPl.C2} and comets \citep{64StGrxx}.

\begin{figure}
\centering
\includegraphics[width=0.6\textwidth]{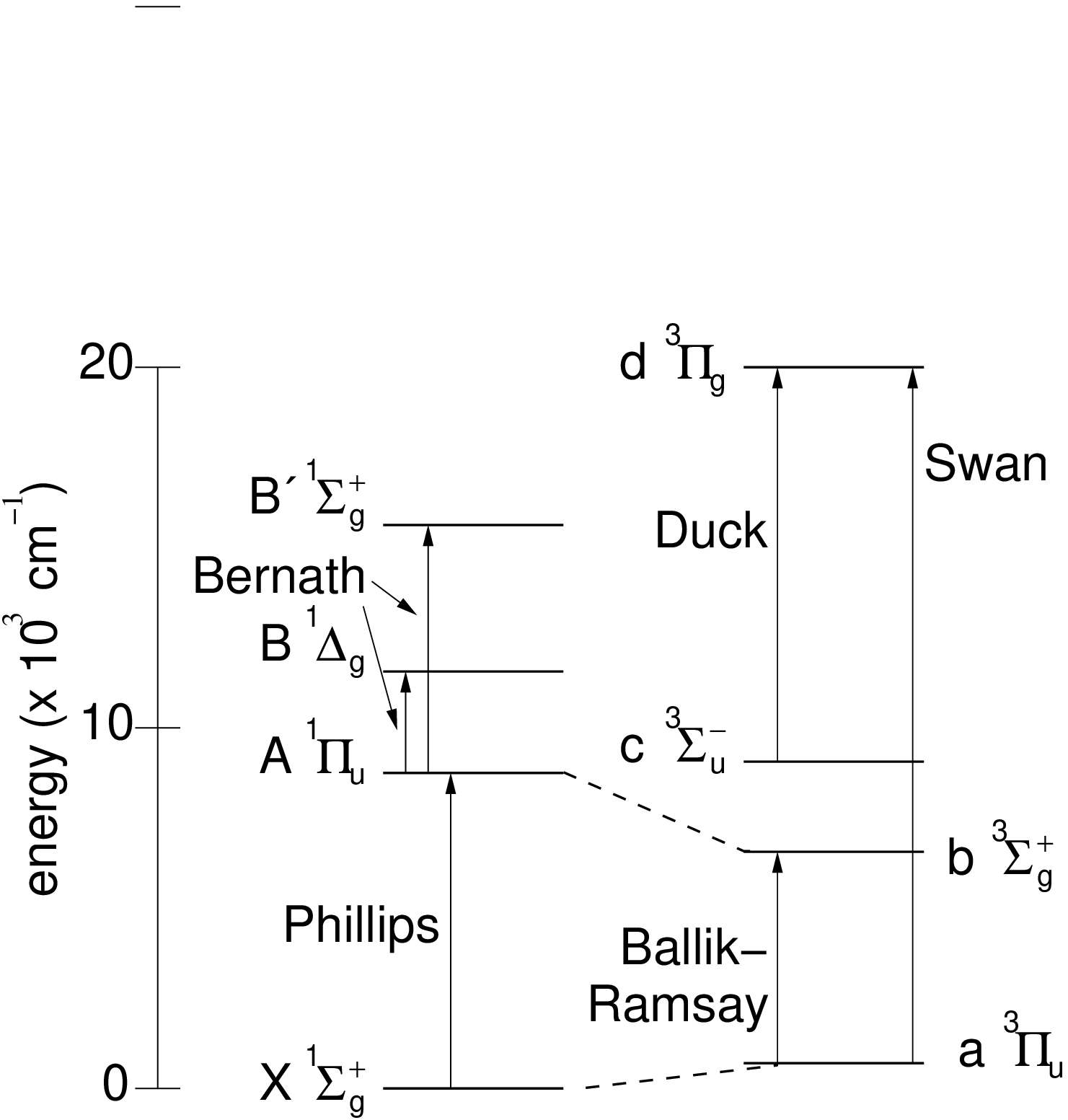}
\caption{Band systems connecting the eight lowest electronic states of C$_2$ considered in this work. The dashed lines represent intercombination  bands.}
\label{f:bands}
\end{figure}

Unusually, astronomical spectra of C$_2$ have been
observed via several different electronic bands; those considered in
this work are summarised in Figure~\ref{f:bands}.
The spectroscopy of C$_2$ is an important tool for stellar
classifications
\citep{41KeMoxx,70Vartya,80Fujita.C2,93Keenan,07GoLaTr,09DeLoAr} and
determining the chemical composition of stars
\citep{71QuGuKu,84LaBrHi,90Goorvitch,96BaWaLa,07GoLaTr,08HaMa.C2,07Green,09ZaAbPl.C2,12IsPaRe,13ScZaPu}
and of the Sun \citep{68Lambert,73GrSaxx,78Lambert,82BrDeGr}.

The Swan bands of C\2\ have
long been known in cometary spectra \citep{1911Meunie.C2}. These
bands are easily detected and have been extensively studied, see
\citet{12RoJeMa}, for example. The Swan bands are useful to estimate the
effective excitation temperatures of C\2, see, for example,
\citet{83LaDaxx} and \citet{95RoMoCl}. Using C\2\ columns
densities, \citet{84NeSpXX} were able to obtain C$_2$/O and C$_2$/CN
ratios for 17 comets. Other observations include works on the cometary spectroscopy of C\2\ are by \citet{64StGrxx,68MaOdxx,73Owen,
74DaLaAr,83LaDaxx,83JoFiLa,84NeSpXX,89GrDiBl,12RoJeMa}.

\cite{64StGrxx} used the (1,0) Swan band to determine
the $^{12}$C/$^{13}$C ratio in  comet
Ikeya  and found it similar to that observed in the solar
system, $^{12}$C/$^{13}$C = 89. The (1,0) band head was also used to
determine the $^{12}$C/$^{13}$C ratio in the comet Tago-Sato-Kosaka
1969 by \citet{73Owen} and the comet Kohoutek by \citet{74DaLaAr}.
\citet{12RoJeMa}  used the (1,0) and (2,1) Swan bands to obtain
 isotope ratios for 2 comets (NEAT and LINEAR), which were
again consistent with the
terrestrial ratio, thus supporting the proposition that comets were
created in our solar system and indicating that the ratio has not changed
significantly since their birth \citep{12RoJeMa}.

Although C$_2$ had long been observed in spectra of cool stars and
comets, the first detection in the interstellar medium (ISM) was made
by \citet{77SoLu} using the C$_2$ Phillips (1,0) band in the
near-infrared spectrum towards the star Cyg OB212.  The $Q(2)$ line of
the Phillips (2,0) band was observed by \citet{78ChLuxx.C2} toward
$\zeta$ Oph, after which \citet{80ChLuBl.C2} observed 9 lines of this
band toward $\zeta$ Per. The (3,0) band was observed by
\citet{86VaBlxx.C2} toward $\zeta$ Oph.  There have followed many
other ISM observations featuring C\2\ spectra \citep{79Hobbs,81Hobbs,82HoCaxx,83HoBlDiH,84vaZexx,86VaBlxx.C2,
  88BlDixx,89FeHuxx.C2,06SnMcxx,07SoWeThYo,10WeRoLi,12HuShFe}.

\citet{86LeRo} suggested that the \astate\ -- \X\ intercombination
band might be observable in the ISM. So far such
lines have yet to be observed in the ISM and attempts to observe them
in a comet also failed \citep{98RoLaMoCl}. However, this band has
recently been detected in the laboratory \citep{15ChKaBeTa.C2},
allowing a precise determination of the singlet-triplet separation \citep{jt637}. Our line list provides accurate wavelengths and transition intensities for
these lines based on rovibronic mixing between states.

On Earth C$_2$ is abundant in flames, explosions, combustion sources,
electrical hydrocarbon discharges and photolysis processes
\citep{11Nemes}.  The C\2\ spectrum (especially the Swan band
$d$--$a$) is commonly used to monitor carbon-based plasmas including
industrial applications; see, for example, \citet{07JoNeCa,13AlHaHa,17BaChRy} and references therein.

Although numerous transition bands have been studied experimentally,
the accuracy of the line positions has been considerably improved in recent
years by application of jet expansion, modern lasers and Fourier
transform techniques. Here we  focus on the most
accurate measured transitions, which involve the first eight lowest electronic
states of C$_2$, see Figure~\ref{f:bands}. A summary of experimental work on the bands
linking these states is presented below
through the Phillips, Swan, Ballik-Ramsay, Bernath and Duck
systems. \citet{jt637} recently undertook a comprehensive assessment of high-resolution
laboratory studies of C$_2$ spectra; they derived empirical energy
levels using the MARVEL (measured active rotation vibration energy levels)
procedure which are used extensively  in the present work.

There have been extensive theoretical studies involving the eight lowest
electronic states of C$_2$;
here we discuss only the most recent
works.
Because of the near degeneracies among the
electron configurations along the whole range of internuclear separations,
the potential energy curves (PECs) lie
very close together, even near the equilibrium geometry,
and several PECs undergo avoided crossings. This means that
traditional single-reference methods
are unable to provide quantitatively acceptable
results for the functions dependent upon the interatomic distance
\citep{04AbSherr.C2,05ShPixx.C2}.

Systematic high level \ai\ analysis
of the $J=0$ vibrational manifolds including also the
$^{12}$C$^{13}$C and $^{13}$C$_2$ isotopologues was performed
for the \A\ and \X\ states by \citet{11XiDeJi.C2},
who computed
PECs of C\2\ at the  multi-reference configuration interaction (MRCI) \citep{88WeKnxx.ai} level of theory in conjunction with the aug-cc-pV6Z
basis set, using complete active space self-consistent field (CASSCF) \citep{80RoTaxx.ai,85WeKnxx.ai} reference
wave functions.

Highly accurate potential energy curves, transition dipole moment functions,
spectroscopic constants, oscillator strengths and radiative
lifetimes were obtained for the Phillips, Swan, Ballik-Ramsay and
Duck  systems
by  \citet{07KoBaSc.C2}
using the CASSCF and subsequent MRCI computational
approach including higher order corrections.
\citet{07ScBaxx.C2} improved the aforementioned
computational methodology  by computation of
MRCI transition dipole moments between
these four systems.
Accurate \ai\ calculations of three PECs of C\2\ at the complete basis set limit were reported by \citet{08Varand}.

\citet{13BrBeScBa.C2} presented an empirical line list for the Swan
system of C\2\ (\dstate--\astate) which included vibrational bands
with $v'=0-10$ and $v''=0-9$, and rotational states with $J$ up to 96,
based on an accurate \ai\ (MRCI) transition dipole moment $d$--$a$
curve.  The opacity database of \citet{11Kurucz.db} contains a C\2\
line list for several electronic bands; Ballik-Ramsay, Swan,
Fox-Herzberg (\estate\ -- \astate) and Phillips.

Experimental lifetimes of specific vibronic states of C\2\ have been
reported by
\citet{69Smithx.C2,76CuEnEr.C2,85BaBeHu,86BaBeBiMe,88NaCoDo.C2,95ErIwam.C2}
and \citet{07KoBaSc.C2}. These observations provide an important test
of any spectroscopic model for the system.  \citet{13BrBeScBa.C2}
reported theoretical lifetimes for the Swan band which were in good
agreement with experimental values.

The ExoMol project aims to provide line lists for all molecules of
importance for the atmospheres of exoplanets and cool stars
\citep{jt528,jt631}.  Given the astrophysical importance of C$_2$ and
the lack of a comprehensive line list for the molecule, it is natural
that C$_2$ should be treated as part of the ExoMol project.  Here we
use the program \duo\ \citep{Duo} to produce line lists for the eight
electronic states (\allstates) of three isotopologues of C$_2$. The
electronic bands connecting these states are summarized in
Fig.~\ref{f:bands}.  The line lists are computed using high level \ai\
transition dipole moments of C$_2$, MRCI/aug-cc-pwCVQZ-DK and
empirical potential energy, spin-orbit, electronic angular momenta, Born-Oppenheimer breakdown, spin-spin, spin-rotation and $\Lambda$-doubling curves (see below for description of the curves taken into account). These empirical curves were obtained by refining \ai\
curves using a recent set of experimentally-derived (MARVEL) term
values of C$_2$ \citep{jt637}.  This methodology has been used for
similar studies as part of the ExoMol project including the diatomic
molecules AlO \citep{jt598}, ScH \citep{jt599}, CaO \citep{jt618}, PO
and PS \citep{jt703}, VO \citet{jt644}, NO \citep{jt686}, NS and SH
\citep{jt725}, SiH \citep{jt711}, and AlH \citep{jt732}.

\section{Theoretical approach}
\label{s:theory}

\subsection{Electronic structure computations}
The presence of spin, orbital and rotational angular moment results in complicated
and extensive couplings between electronic states. How these are treated formally
and their non-perturbative inclusion in the calculation of rovibronic spectra of diatomic molecules
is the subject of a recent topical review by two of us \citep{jt632};
this review provides a detailed, formal description of the various coupling curves considered below.

\begin{figure}
\centering
\includegraphics[width=0.6\textwidth]{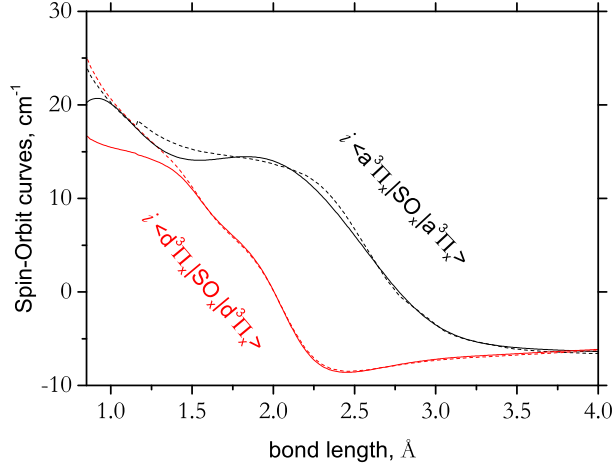}
\caption{Diagonal spin-orbit curves of C$_2$ between the \astate\ and  \dstate. The \ai\ curves are shown using dashed line,  while the refined curves are given by solid lines. The empirical \dstate\ SOC was produced by morphing the \ai\ curve, while the \astate\ SOC was obtained by refining the \ai\ parameters.}
\label{f:SO:0}
\end{figure}

\begin{figure}
\centering
\includegraphics[width=0.6\textwidth]{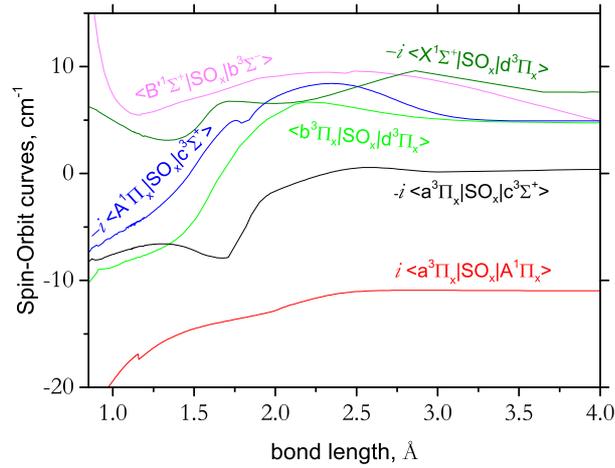}
\caption{\Ai\ $a$--$A$, $a$--$c$,  $B^\prime$--$b$, $b$--$d$, $X$--$d$ and $A$--$c$ spin-orbit curves of C$_2$. These curves were not refined.}
\label{f:SO:1}
\end{figure}

\begin{figure}
\centering
\includegraphics[width=0.6\textwidth]{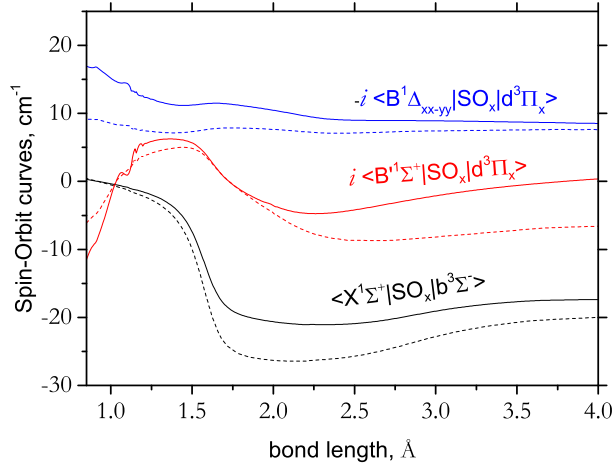}
\caption{Spin-orbit $X$--$b$, $B$--$d$ and $B^\prime$--$d$ curves of C$_2$. The \ai\ curves are shown using dashed line,  while the refined curves are given by solid curves. }
\label{f:SO:2}
\end{figure}

\begin{figure}
\centering
\includegraphics[width=0.6\textwidth]{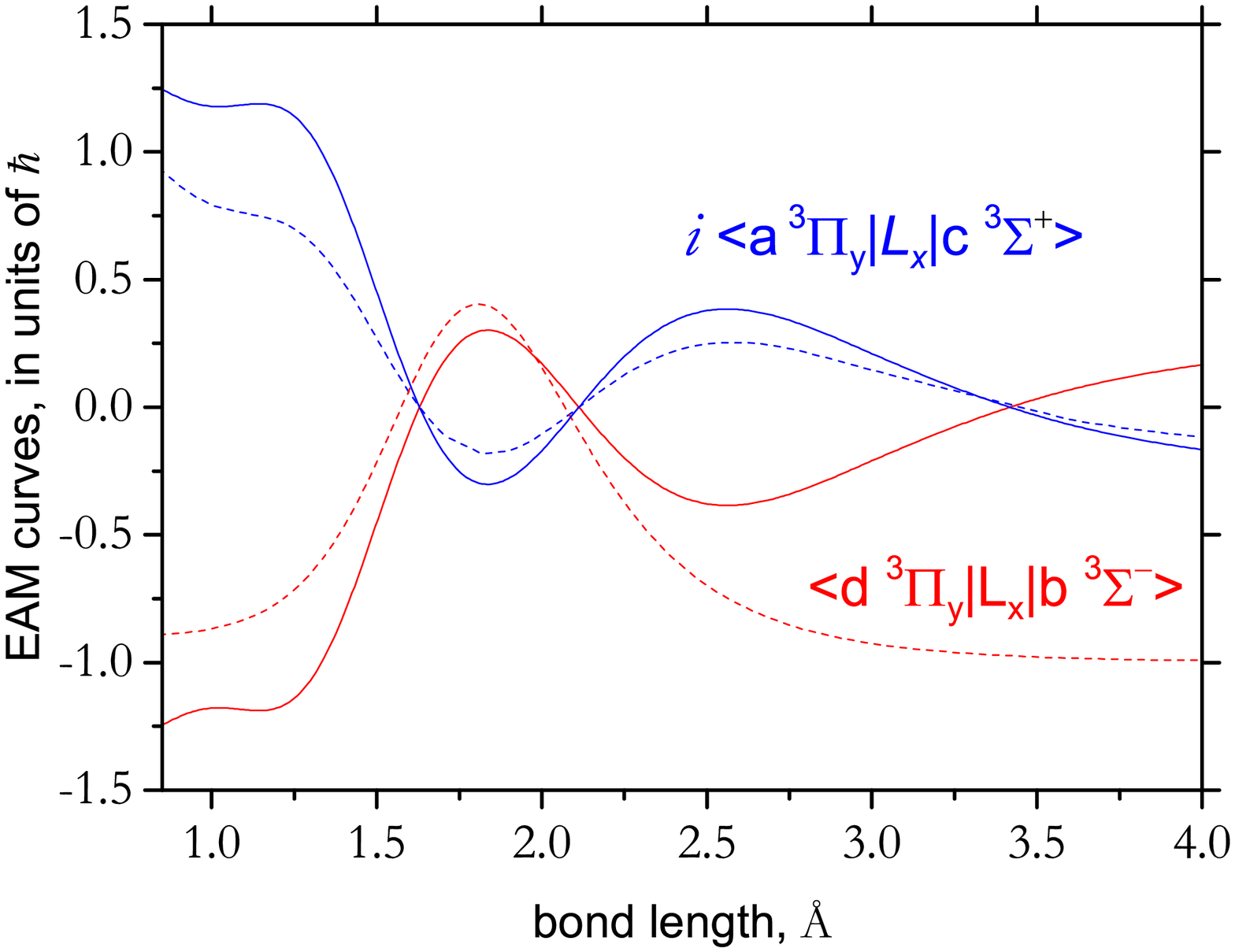}
\caption{Refined $a$--$c$ and $d$--$b$ EAM curves of C$_2$. The \ai\ curves are shown using dashed line,  while the refined curves are given by solid curves. }
\label{f:EAM}
\end{figure}

The PECs,  spin-orbit coupling curves (SOCs), electronic angular momentum curves (EAMCs)
and the transition dipole moment curves (TDMs) were computed
at the MRCI level of theory, using reference wave functions from a
CASSCF with all single and double excitations included,
in conjunction with the augmented correlation-consistent
polarized aug-cc-pwCVQZ-DK Dunning type basis set \citep{89Dunning.ai, 93WoDuxx.ai,02PeDuxx.ai}, plus Duglas-Kroll corrections and core-correlation effects as implemented in MOLPRO \citep{MOLPRO}.
The complete active space is defined by (3,1,1,0,3,1,1,0) in the D$_{2h}$ symmetry group employed by MOLPRO, which corresponds to the  $A_{g}$, $B_{3u}$, $B_{2u}$, $B_{1g}$, $B_{1u}$, $B_{2g}$, $B_{3g}$ and $A_u$ irreducible representations of this group, respectively. The initial grid included about 400 points ranging from 0.7 to 10~\AA, however some geometries close to the curve crossings did not converge and were then excluded.  Some of the \ai\ curves are shown in Figs.~\ref{f:PECs} --
\ref{f:DMC}. Our $a$--$d$ transition dipole moment curve compares well
with that computed by \citet{13BrBeScBa.C2} who used it to produce their C\2\
line list for the Swan system.

\subsection{Solution of the rovibronic problem}

We used the program \duo\ \citep{Duo} to solve the fully coupled
Schr\"{o}dinger equation for eight lowest electronic states of C\2,
single and triplet: \allstates. The vibrational basis set was
constructed by solving eight uncoupled Schr\"{o}ninger equations using
the sinc~DVR method based on the grid of equidistant 401 points
covering the bond lengths between 0.85 and 4 \AA. The vibrational
basis sets sizes were 60, 30, 30, 30, 40, 40, 30 and 30 for
\allstates, respectively.

\duo\ employs Hund's case a formalism:
rotational and spin basis set functions are the
spherical harmonics $\ket{J,\Omega}$ and $\ket{S,\Sigma}$, respectively.
For the nuclear-motion step of the calculation, the
electronic basis functions $\ket{{\rm State},\Lambda}$ are defined
implicitly by the matrix elements of the SO, EAM coupling and TDM as
computed by MOLPRO. Note that the couplings and TDMs had to be made
phase-consistent \citep{jt589} and
transformed to the symmetrized $\Lambda$-representation, see
\citet{Duo}.

\subsection{Potential energy curves}

Our PECs are fully empirical (reconstructed through the fit to the experimental data). To represent the potential energy curves the following two types of functions
were used.

For the simpler PECs that do not exhibit avoided crossing (\A, \B, \astate,
\bstate\ and \cstate) we used the extended Morse oscillator (EMO) functions
\citep{EMO} for both \ai\ and refined PECs.

In this case a PEC is given by
\begin{equation}\label{e:EMO}
V(r)=V_{\rm e}\;\;+\;\;(A_{\rm e} - V_{\rm
e})\left[1\;\;-\;\;\exp\left(-\sum_{k=0}^{N} B_{k}\xi_p^{k}(r-r_{\rm e})
\right)\right]^2,
\end{equation}
where $A_{\rm e} - V_{\rm e}$ is the dissociation energy, $r_{\rm e}$ is an equilibrium distance of the PEC, and $\xi_p$ is the \v{S}urkus variable given by:
\begin{equation}
\label{e:surkus:2}
\xi_p= \frac{r^{p}-r^{p}_{\rm e}}{r^{p}+r^{p}_{\rm e }}.
\end{equation}
The corresponding expansion parameters are obtained by fitting to the empirical (MARVEL) energies from \citet{jt637}.

\begin{figure}
\centering
\includegraphics[width=0.49\textwidth]{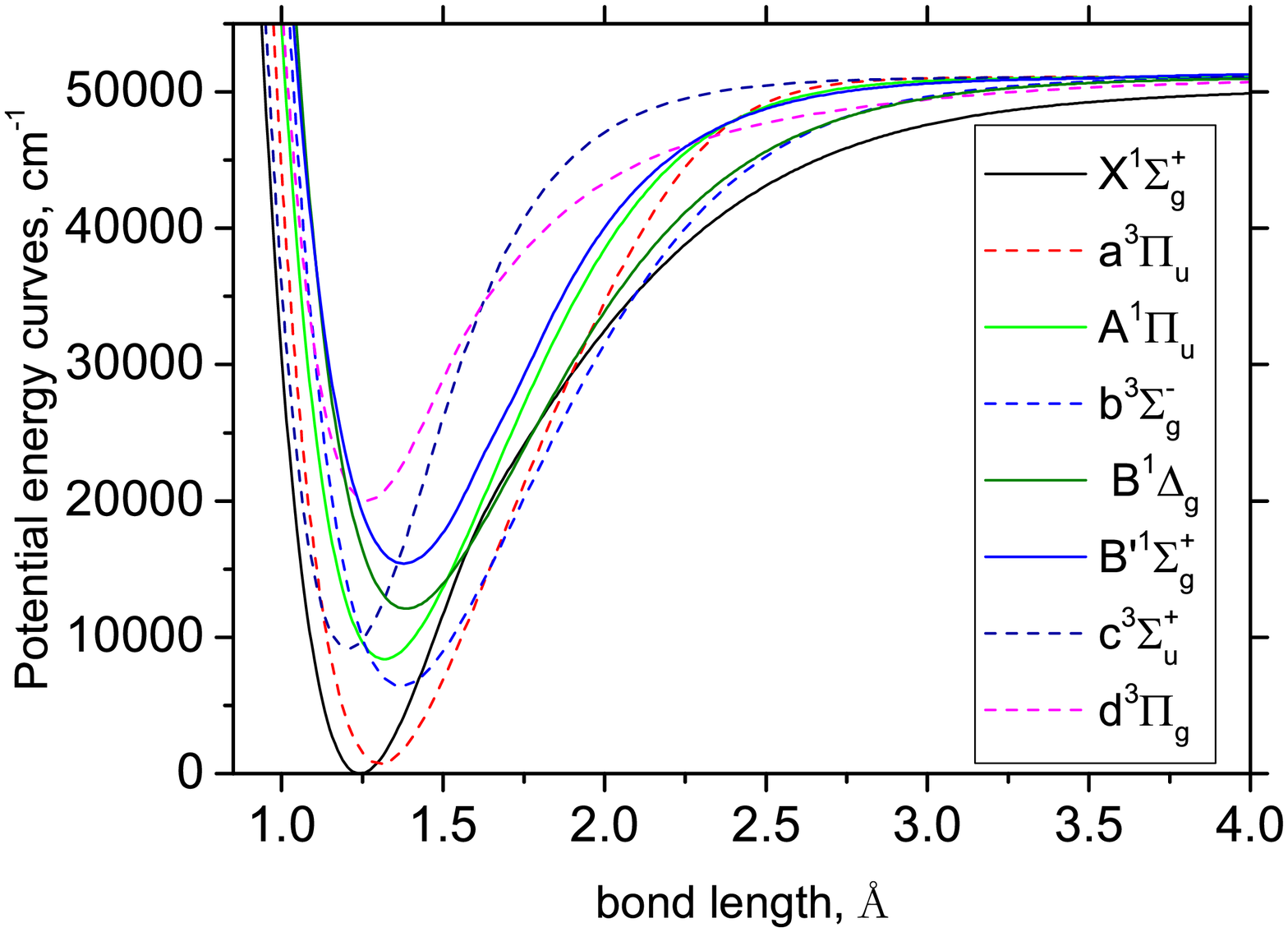}
\includegraphics[width=0.49\textwidth]{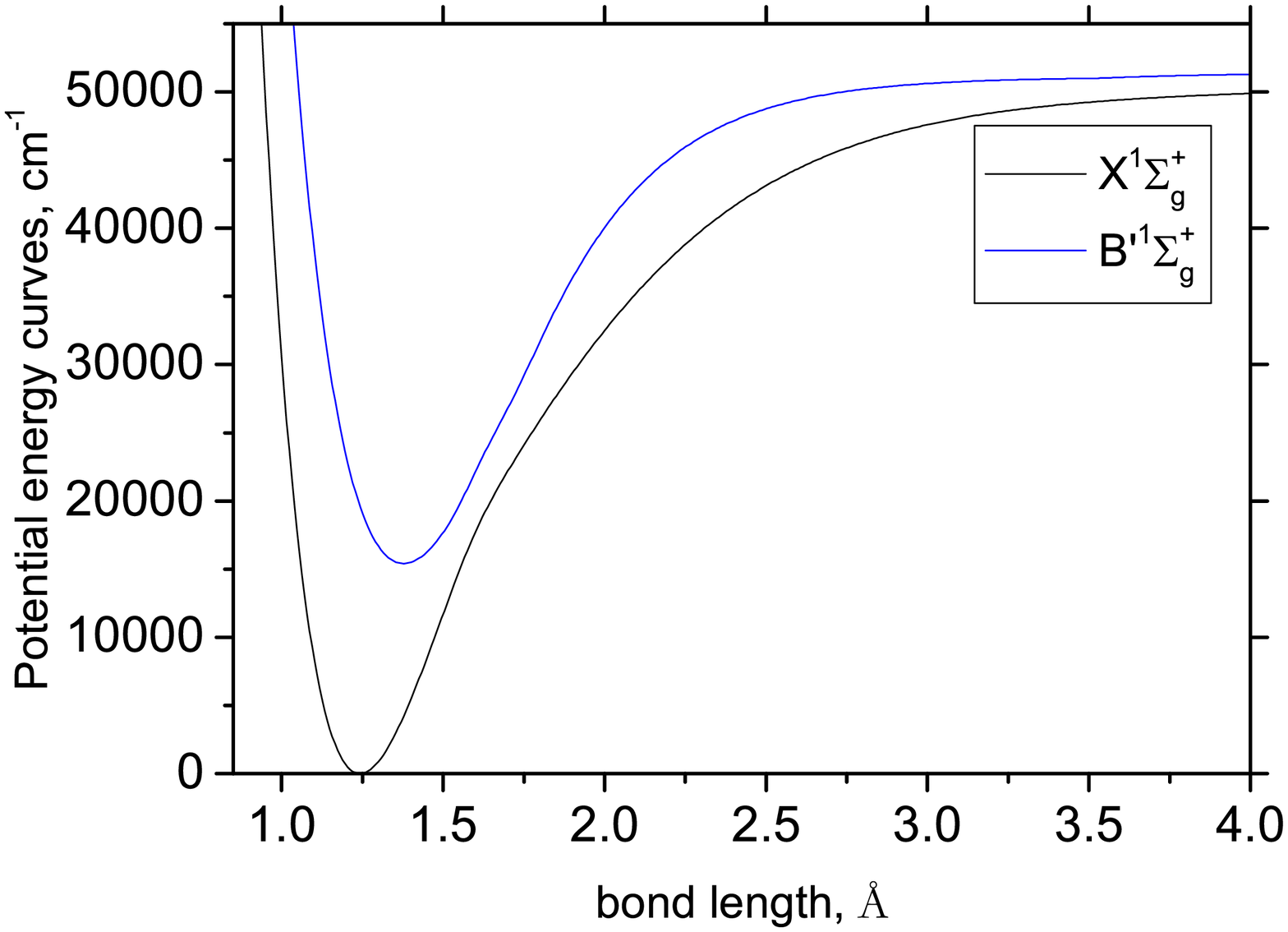}
\caption{The eight refined potential energy curves of C$_2$ (left) and
the avoided crossing between the \X\ and \Bp\ states in
the adiabatic representation  (right). }
\label{f:PECs}
\end{figure}

For the three states with avoided crossing, \X, \Bp\ and \dstate\
(see Fig.~\ref{f:PECs}) a diabatic representation of two coupled EMO PECs was used. In this representation the PEC is obtained as a root of a characteristic  $2\times 2$ diabatic matrix
\begin{equation}\label{e:V1W/WV2}
\bf{A} = \left(
\begin{array}{cc}
  V_1(r) & W(r) \\
  W(r) & V_2(r)
\end{array}
\right),
\end{equation}
where $V_1(r)$ and $V_2(r)$ are given by the EMO potential function in Eq.~\eqref{e:EMO}. The coupling function $W(r)$ is given by
\begin{equation}\label{e:W(r)}
W(r) = W_0 + \frac{\sum_{i\ge 0} w_{i} (r-r_{\rm cr})^i  }{\cosh[\beta (r-r_{\rm cr})]},
\end{equation}
where $r_{\rm cr}$ is a crossing point. The two eigenvalues  of the matrix $\bf{A}$ are given by
\begin{eqnarray}
  V_{\rm low}(r) &=& \frac{V_1(r)+V_2(r)}{2}-\frac{\sqrt{[V_1(r)-V_2(r)]^2+4 \, W^2(r)}}{2}, \\
  V_{\rm upp}(r) &=& \frac{V_1(r)+V_2(r)}{2}+\frac{\sqrt{[V_1(r)-V_2(r)]^2+4 \, W^2(r)}}{2}.
\end{eqnarray}

For each pair of states, only one component is taken, $V_{\rm low}$
for \X\ and \dstate\ and $V_{\rm upp}$ for \Bp, and the other component is
ignored. For example, the coupled $X$--$B^\prime$ system is treated as
two independent diabatic systems in Eq.~\eqref{e:V1W/WV2}, as we could
not obtain a consistent model with only one pair of the $X$ and
$B^\prime$ curves. In case of the \X\ state, the upper component,
formally representing the \Bp\ state, is only used as a dummy PEC. The
actual \Bp\ PEC is taken as the upper component with different $V_{\rm
  low}$. The latter is also a dummy PEC and disregarded from the rest
of the calculations. In this decoupled way we could achieve a more
stable fit.

The expansion parameters, including the corresponding equilibrium bond
lengths $r_\rme$ appearing in Eqs.~\eqref{e:EMO}--\eqref{e:W(r)} are
obtained by fitting to the experimentally-derived energies. The
dissociation asymptote $A_\rme$ in all cases was first varied and then
fixed the value 50937.91~\cm\ (6.315 eV) for all but the \dstate\ PEC,
for which it was refined to obtain 62826.57~\cm\ (7.789~eV) for better
accuracy. To compare, the experimental value of $D_0$ = 6.30 $\pm
0.02$~eV ($D_{\rme}$ $\sim$ 6.41~eV) was determined by
\citet{91UrBaJa}. The best \ai\ values of $D_\rme$ of C\2\ from the
literature include 6.197~eV by \citet{00FeSoxx} and 6.381~eV by
\citet{08Varand}.  The lowest asymptote $A_\rme$ correlates with the
$^3P$+$^3P$ limit \citep{92Martin.C2}, while the next is the
$^3P$+$^1D$ limit (+1.26~eV).
Our zero-point-value is 924.02 \cm.



The effect of the avoiding crossings on the shape of the \X, \Bp\ and
\dstate\ PECs is illustrated in Fig.~\ref{f:PECs}. It is clear that
simple one-curve expansions would be problematic for these states.
This effect has been studied in detail by \cite{08Varand,09Varand.C2}.

\subsection{Couplings}

In the refinement of the SO and EAM coupling we use the \ai\ curves, which are `morphed' at the \ai\ grid points using the following  expansion:
\begin{equation}
\label{e:bob}
F(r)=\sum^{N}_{k=0}B_{k}\, z^{k} (1-\xi_p) + \xi_p\, B_{\infty},
\end{equation}
where $z$ is either taken as the \v{S}urkus variable $z=\xi_p$  or a
damped-coordinate given by:
\begin{equation}\label{e:damp}
z = (r-r_{\rm ref})\, e^{-\beta_2 (r-r_{\rm ref})^2-\beta_4 (r - r_{\rm ref})^4},
\end{equation}
see also \citet{jt703} and \citet{jt711}.  Here $r_{\rm ref}$ is a
reference position equal to $r_{\rm e}$ by default and $\beta_2$ and
$\beta_4$ are damping factors.  When used for morphing, the parameter
$B_{\infty}$ is usually fixed to 1. The $B_{\infty}$ parameters should
in principle correspond to the atomic limit of the corresponding
couplings, however we have not attempted to apply any such constraints. Due
to very steep character of the potential energy curves, the long-range
part of the coupling curves has no impact on the states we consider.

Some of the coupling curves have complex shapes due to, for example,
avoiding crossings. This complexity is assumed to be covered by the
morphing procedure, as morphed curves should inherit the shape of the
parent function.


\begin{figure}
\centering
\includegraphics[width=0.45\textwidth]{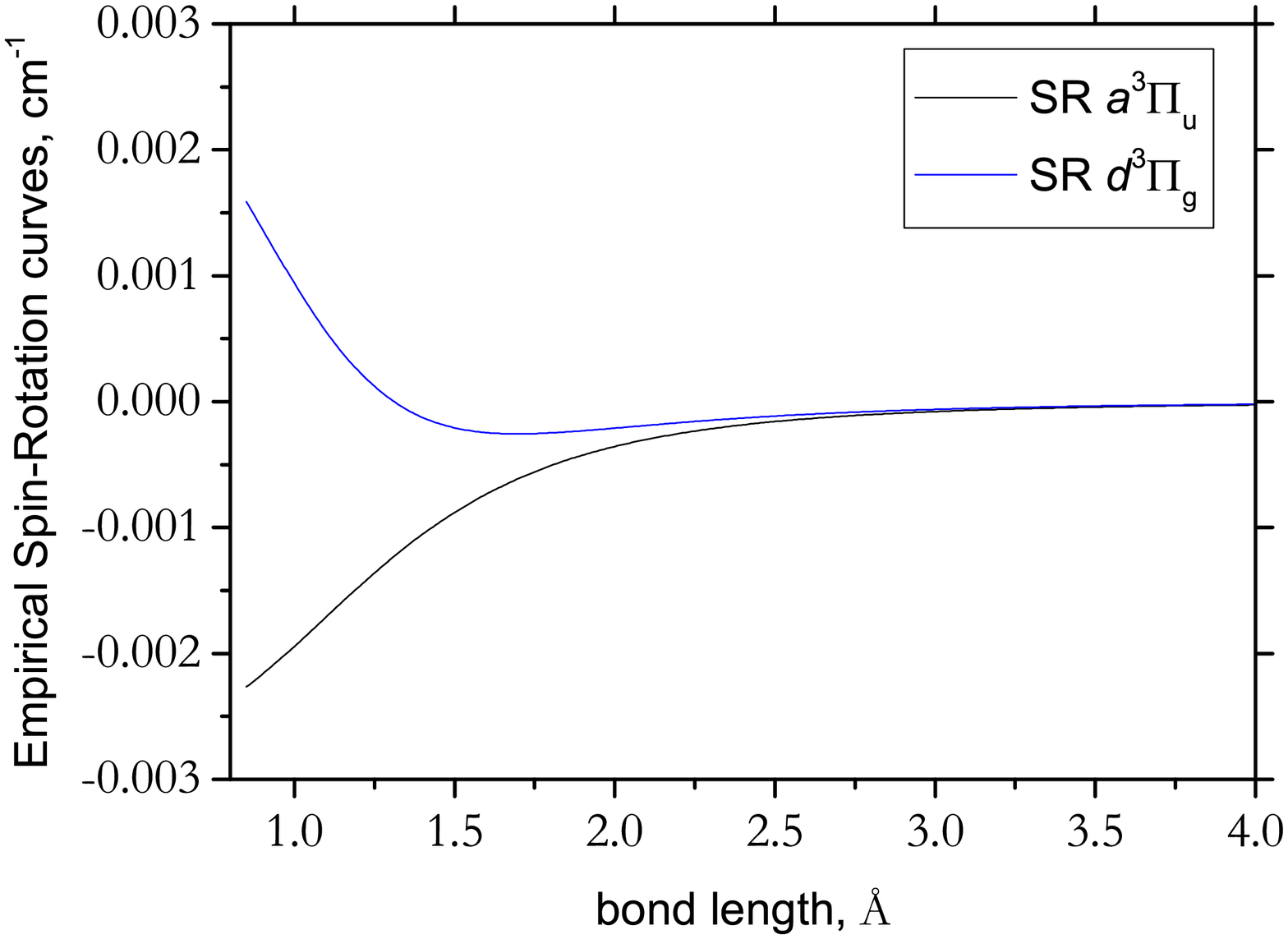}
\includegraphics[width=0.45\textwidth]{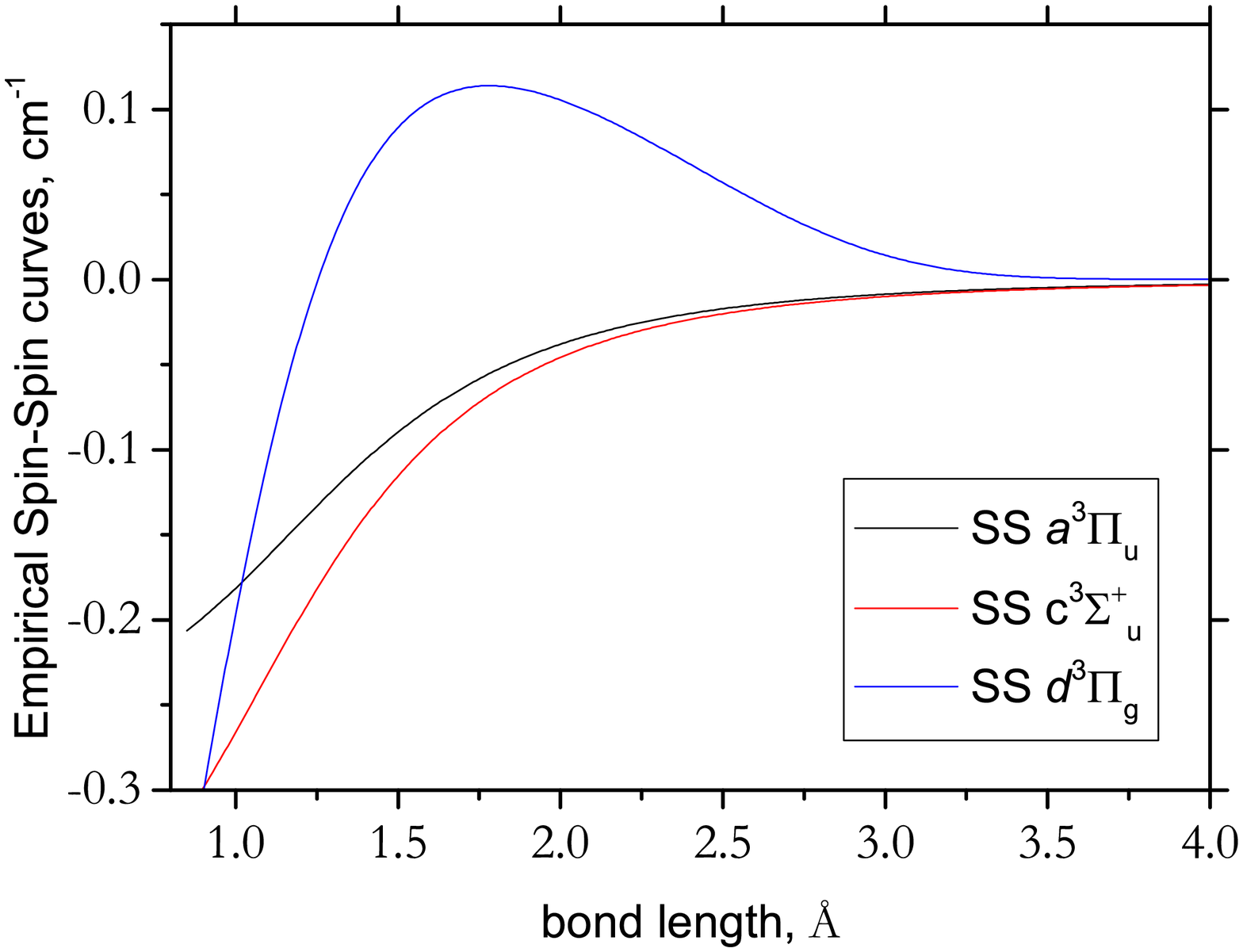}
\caption{Empirical curves of C$_2$: spin-rotational (left) and spin-spin (right) curves representing the states \astate, \cstate\ and \dstate.}
\label{f:SS:SR}
\end{figure}

The spin-spin and spin-rotational couplings were introduced for the states \astate, \cstate\ and \dstate\ and also modelled using the expansion given by Eq.~\eqref{e:bob}. The final curves, which are fully empirical, are shown in Fig.~\ref{f:SS:SR}.

\begin{figure}
\centering
\includegraphics[width=0.45\textwidth]{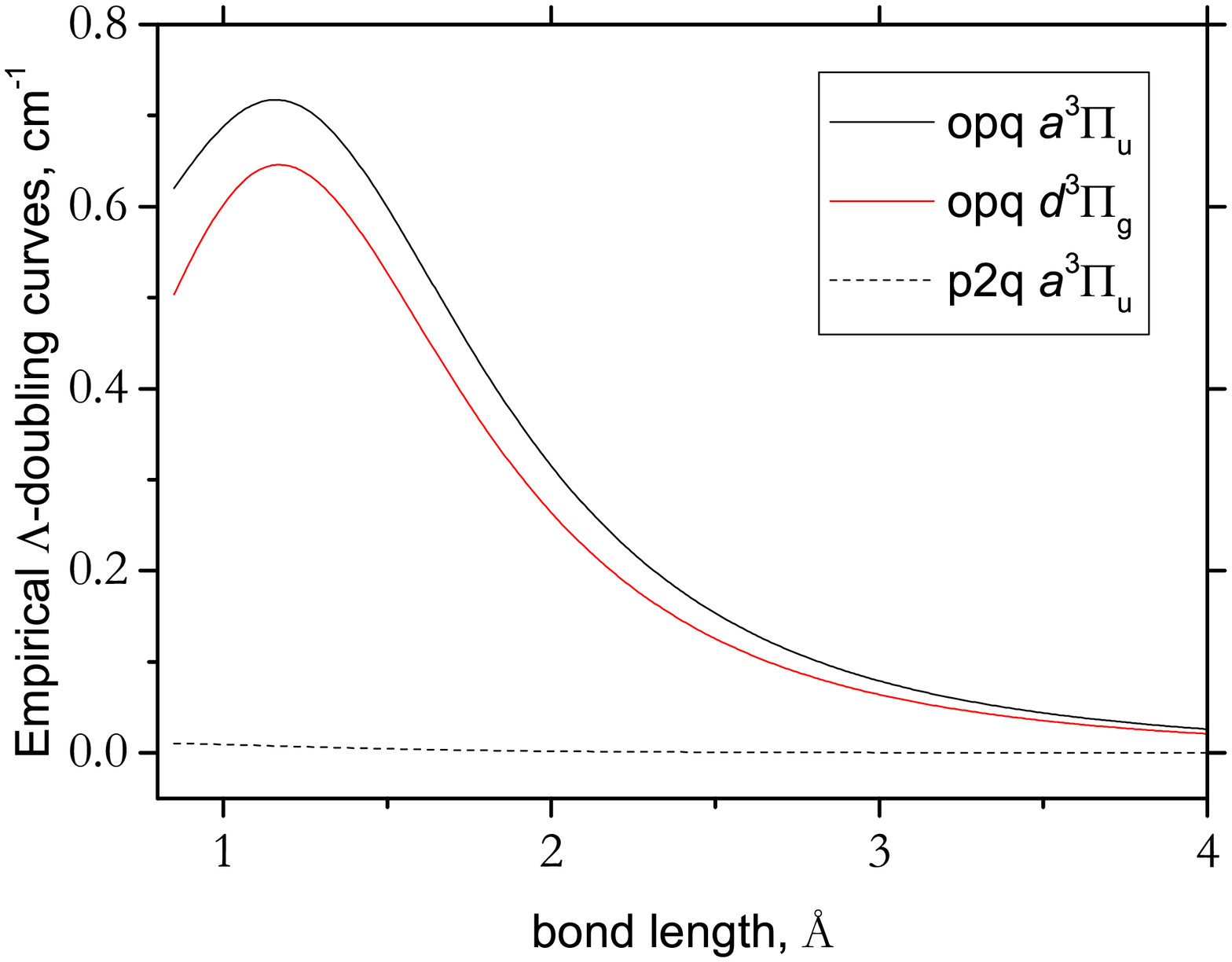}
\includegraphics[width=0.45\textwidth]{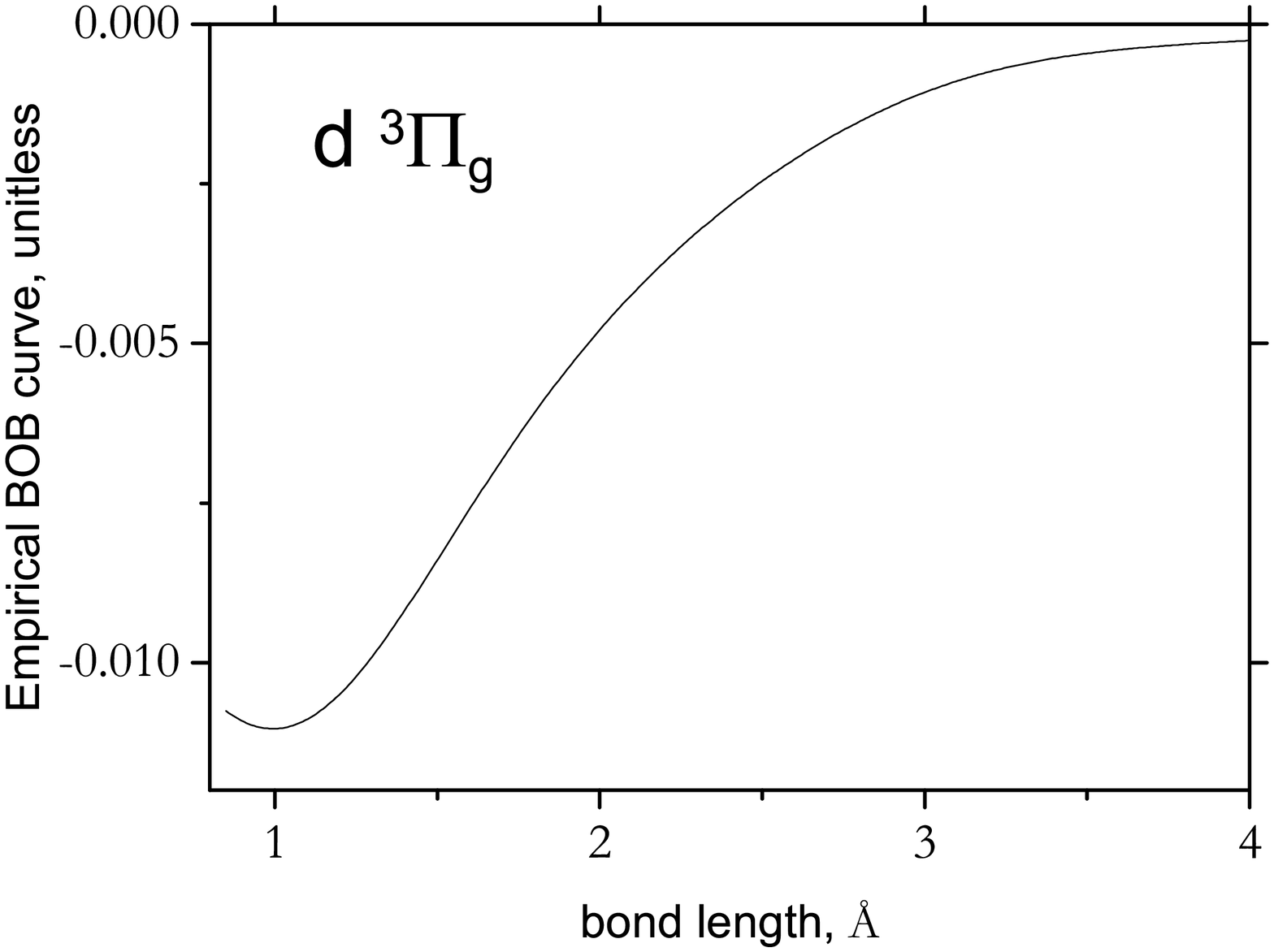}
\caption{Empirical the $\Lambda$-doubling curves used for \astate\ and \dstate (left) and a Born-Oppenheimer breakdown curve used for \dstate\ (right) of C$_2$.}
\label{f:BOB}
\end{figure}

The  $\Lambda$-doubling effects in \astate\ and \dstate\ were obtained empirically  using  effective $\Lambda$-doubling functions, the ($o$+$p$+$q$) and ($p$+2$q$) coupling operators \citep{79BrMexx.methods} as given by:
\begin{eqnarray}
\label{e:Hlambda-doubling}
\hat{H}_{opq} &=& \frac{1}{2} \alpha_{\rm opq}^{\rm LD }(r)  \left( \hat{S}_+^2+\hat{S}_{-}^{2}  \right), \\
\hat{H}_{\rm p2q} &=&
- \frac{1}{2} \alpha_{p2q}^{\rm LD }(r)  \left( \hat{J}_{+}\hat{S}_{+} + \hat{J}_{-}\hat{S}_{-} \right).
\end{eqnarray}
The latter operator is limited to linear $\hat{J}$-dependence, which is
justified for the heavy molecule like C\2. In this case for
$\alpha_{p2q}^{\rm LD }(r)$ and $\alpha_{\rm opq}^{\rm LD }(r)$ we use the \v{S}urkus-type expansion as in
Eq.~\eqref{e:bob}. The empirical $\Lambda$-doubling curves of C\2\ are
shown in Fig.~\ref{f:BOB}. We used these couplings to improve the fit
for the states \astate\ and \dstate.

To allow for rotational Born-Oppenheimer breakdown (BOB) effects
\citep{Level}, the vibrational kinetic energy operator for each
electronic state was extended by
\begin{equation}
\label{e:dist}
-\frac{\hbar^2}{2 \mu r^2} \to -\frac{\hbar^2}{2 \mu r^2} \left( 1 + g^{\rm BOB}(r)\right),
\end{equation}
where the unitless BOB functions $g^{\rm BOB}$ are represented by the polynomial
\begin{equation}
 g^{\rm BOB}(r)= \left[ (1 - \xi_p) \sum_{k\ge 0}^{N_{T}} A_k \xi_p^k  + \xi_p A_{\rm \infty} \right],
\label{eq:bobleroy}
\end{equation}
where $\xi_p$ as the \u{S}urkus variable and $p$, $A_k$ and  $A_{\rm \infty}$ are adjustable parameters. This representation was used for the \dstate\ state only, which appeared to be most difficult to fit.

\subsection{Dipole moment curves}

\begin{figure}
\centering
\includegraphics[width=0.45\textwidth]{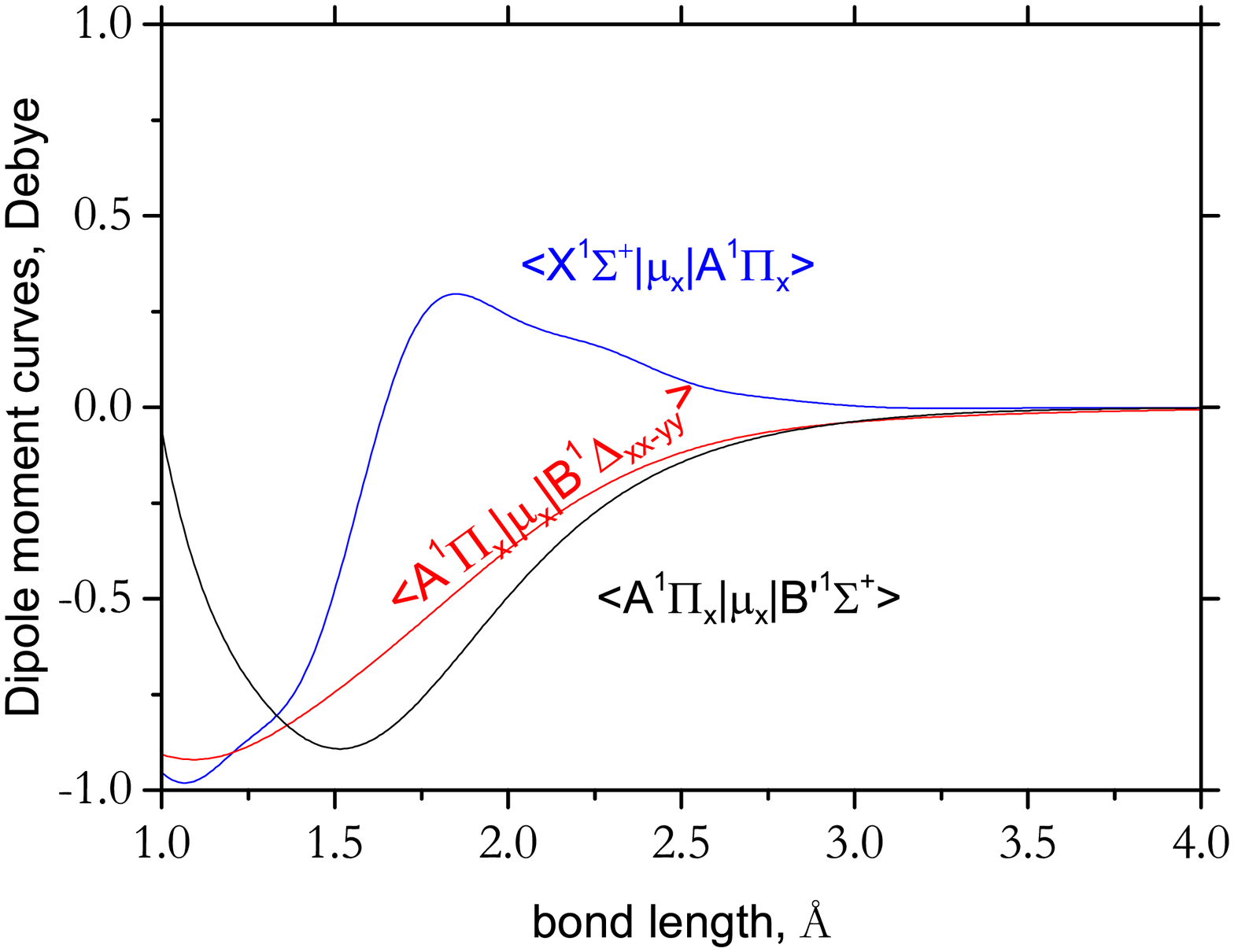}
\includegraphics[width=0.45\textwidth]{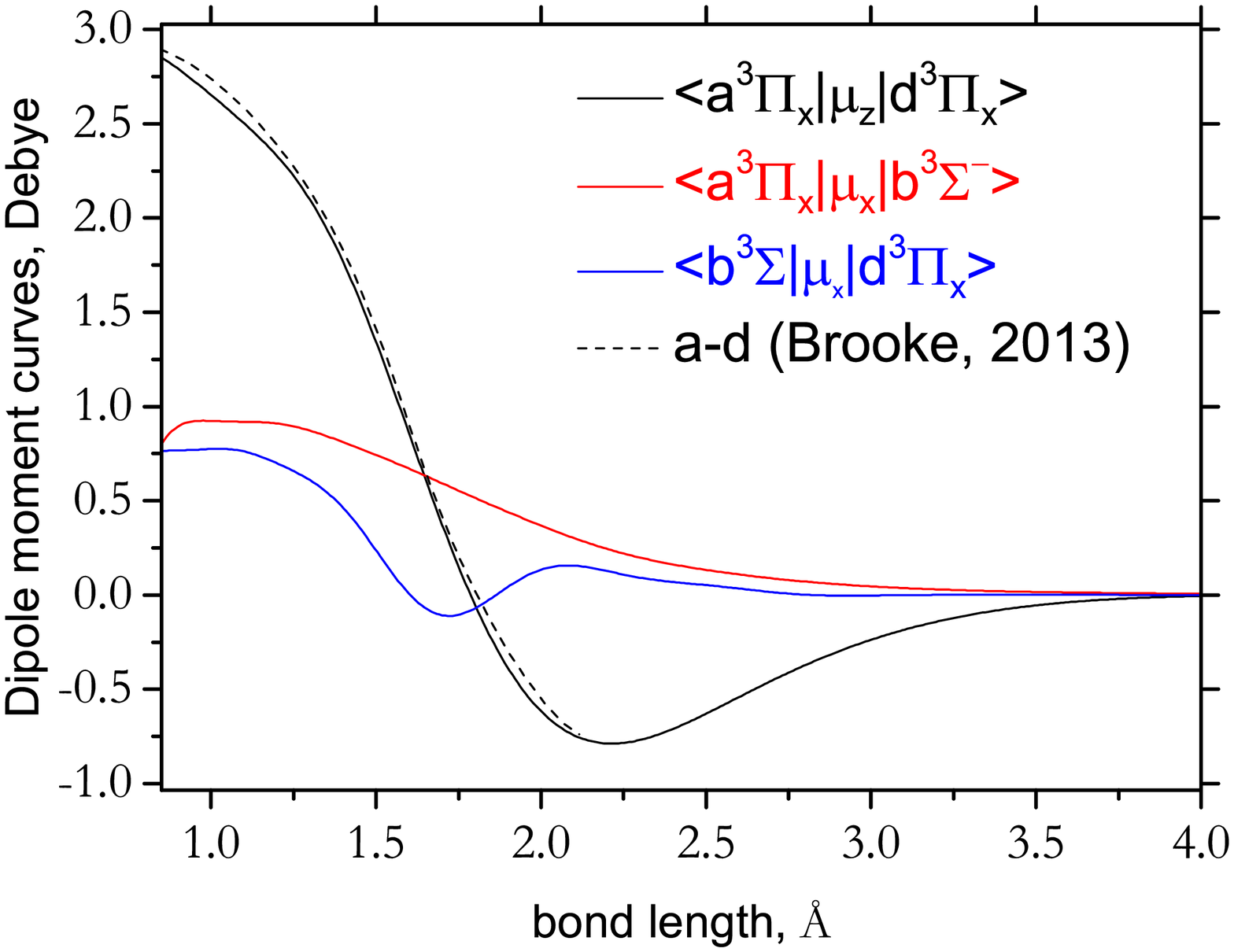}
\caption{Dipole moment curves of C$_2$ between the singlet states $X$--$A$, $A$--$B$ and $A$--$B^\prime$ (left)
and between triplet states $a$--$d$, $a$--$b$ and $b$--$d$. The \ai\ $a$--$d$ TDMC used by \protect\citet{13BrBeScBa.C2}
is also shown. }
\label{f:DMC}
\end{figure}

The electronic dipole pure rotation and rotation-vibration transitions
are forbidden for the homonuclear molecule C\2, so are the transitions
between electronic states with $\Delta \Lambda > 1$ or $\Delta \Lambda = 0$ for $\Sigma$ states.
There are six (electric-dipole) allowed electronic bands between
lowest eight electronic states of C\2 shown in Fig.~\ref{f:bands}. The
corresponding electronic transition dipole moments are shown in
Fig.~\ref{f:DMC}. These \ai\ TDMCs  were represented analytically
using the damped-$z$ expansion in Eq.~\eqref{e:bob}.  This was done in
order to reduce the numerical noise in the calculated intensities for
high overtones, see recent recommendations by \citet{16MeMeSt}. The
corresponding expansion parameters as well as their grid
representations can be found in the \duo\ input files provided as
supplementary data.

\section{Refinement}

In the refinements we used the experimentally-derived energies
obtained by \citet{jt637} using the MARVEL approach. These energies
were based on a comprehensive set of experimental frequencies
collected from a large number of sources which are listed in
Table~\ref{t:sources}. Some statistics about the experimental energies is
shown in Table~\ref{t:stat}. For full details of the MARVEL procedure
as well as descriptions of the experimental data see
\citet{jt637}.

\begin{table}
  \centering
  \caption{Experimental sources used to generate the MARVEL energies of C\2\ \protect\citep{jt637}. }\label{t:sources}
  \begin{tabular}{ll}
\toprule
\multicolumn{1}{c}{System} & \multicolumn{1}{c}{Source} \\
\hline
Phillips \A$-$\X &  \citet{15ChKaBeTa.C2,88DaAbPh.C2,88DoNiBea.C2,77ChMaMa.C2} \\
& \citet{63BaRab.C2,06PeSi,04ChYeWoLi,13NaEn.C2},
\\
Bernath \B\ $-$ \A &
\citet{88DoNiBe.C2,16ChKaBeTa.C2} \\
Bernath \Bp $-$\A & \citet{88DoNiBe.C2} \\
Freymark \Estate$-$\A &
 \citet{97SoBlLiXu,51Freymark.C2} \\
Ballik--Ramsay \bstate$-$\astate &
\citet{15ChKaBeTa.C2,85YaCuMeCa,85RoWaMiVe.C2,79AmChMa.C2}\\
& \citet{88DaAbSa.C2,06PeSi,11BoSyKnGe.C2} \\
Swan    \dstate$-$\astate & \citet{13NaEn.C2,14NaEn.C2,13BoSyKnGe,13YeChWa.C2}\\ & \citet{07TaHiAm.C2,48Phillips.C2,02TaAm.C2,03KaYaGuYu}\\
& \citet{85CuSa,94PrBe.C2,83Amiot,99LlEw} \\
Fox--Herzberg  e~$^{3}\Pi_{\rm g}-$\astate & \citet{86HaWi,49Phillips.C2,98BrHaKoCr} \\
Duck \dstate$-$\cstate &  \citet{13ChYeWa.C2,13NaEn.C2,14NaEn.C2,07JoNaRe.C2}\\
Krechkivska--Schmidt  $4^3\Pi_{\rm g}-$ $a$  &\citet{15KrBaTrNa} \\
Interconnection \astate$-$\X &  \citet{15ChKaBeTa.C2} \\
\A$-$\bstate &  \citet{15ChKaBeTa.C2} \\ $1~^{5}\Pi_{\rm g} -$ \astate &
\citet{11BoSyKnGe.C2} \\
Radi--Bornhauser $1^5\Pi_{\rm u}-1^5\Pi_{\rm g}$ &
\citet{15BoMaGo} \\
\bottomrule
\end{tabular}

\end{table}

The final model comprises 89 parameters appearing in the expansions
from Eqs.~(\ref{e:EMO},\ref{e:W(r)},\ref{e:bob}) obtained by fitting
to 4900 MARVEL energy term values of $^{12}$C\2\ using \Duo. The
robust weighting method of \citet{03Watson.methods} was used to adjust
the fitting weights. During the fit, in order to avoid unphysically
large distortions, the SOC and EAMS curves were constrained to the
\ai\ shapes using the simultaneous fit approach \citep{03YuCaJe.PH3}.
The MARVEL energies were correlated to the theoretical values using
the \Duo\ assignment procedure, which is based on the largest basis
function contribution \citep{Duo}.  One of the main difficulties in
controlling the correspondence between the theoretical (\Duo) and
experimental energies in case of such a complex, strongly coupled
systems is that the relative order of the computed energies can change
during the fit, in this case automatic assignment is especially
helpful.  For some C\2\ resonance states it was necessary to use also
the second largest coefficients to resolve possible ambiguities.
However, even this did not fully prevent accidental re-ordering of
states, especially the assignment of the different $\Omega$ components
of the triplet $a$ and $d$ states appeared to be very unstable and
difficult to control. In such cases, as the final resort for
preventing disastrous fitting effects, states exhibiting too large
errors (typically $>$ 8 \cm) were removed from the fit, which,
together with the second-largest-coefficient assignment feature are
new implementations in \Duo.  One of the artifacts of the
largest-contribution assignment is that it can fail for the
vibrational quantum numbers at high rotational excitations $J$.  Our vibrational basis functions, generated as
eigensolutions of the pure, uncoupled $J=0$, Schr\"{o}dinger equations,
become less efficient for high $J$ ($>50$). This is because of the
centrifugal distortion term in the Hamiltonian, which becomes large
and thus distorts the effective shape of the interaction potential
substantially. The rovibronic eigenfunctions in this case are a
complicated mixture of the $J=0$ vibrational contributions with no distinct
large contributions. The typical situation at high rotational
excitations ($J>50$) is that the rovibronic eigenfunctions consist of
a large number of similar vibrational contributions, which make the
largest-coefficient assignment of the quantum number $v$ meaningless.
Therefore, we treated the vibrational assignment differently: for each
value of $J$, the vibrational quantum number $v$ was assigned by
simply counting states of the same electronic term and
$\Omega$-component. This is another new feature in \Duo\ implemented
in order to improve the vibrational assignment of the states with high
values $J$.

\begin{table}
  \centering
  \caption{Some statistics of the experimental term values of C$_2$ used in this work to refine the model. }\label{t:stat}
  \begin{tabular}{lrr}
\toprule
State    &  $v_{\rm max}$  &  $J_{\rm max}$   \\
    \hline
\X       &               9 &             74   \\
\A       &              16 &             75   \\
\B       &               8 &             50   \\
\Bp      &               3 &             32   \\
\astate  &              14 &             86   \\
\bstate  &               8 &             74   \\
\cstate  &               7 &             24   \\
\dstate  &              12 &             87   \\
\bottomrule
\end{tabular}

\end{table}

It should be noted that not all experimentally-derived MARVEL energies in our
fitting set are supported by multiple transitions and are therefore not equally
reliable. Furthermore, in some cases there is no agreement between different
experimental sources of C\2\ spectra. A particular example is the  $d$--$a$
study by \citet{11BoSyKnGe.C2} who pointed out a 1--2~\cm\ discrepancies with
values from a previous study by \citet{07TaHiAm.C2} for the $P_1(5)$ and
$R_1(5)$ lines $(v'=6,v''=5)$. We obtained similar residuals for these two
transitions.

\begin{figure}
\centering
\includegraphics[width=0.6\textwidth]{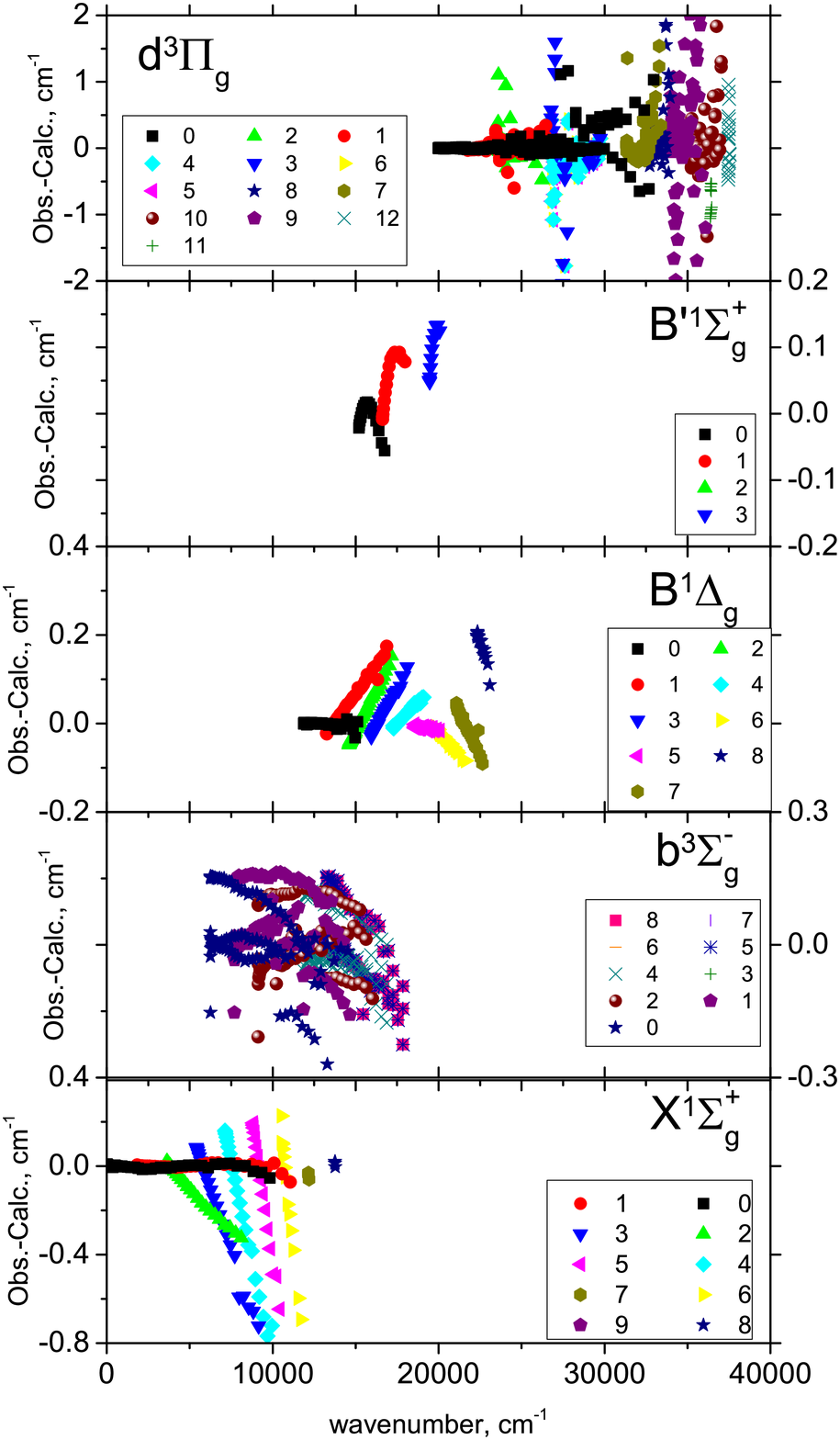}
\caption{Obs.$-$calc. residuals for individual vibronic bands of C$_2$: $g$-states. }
\label{f:obs-calc}
\end{figure}

\begin{figure}
\centering
\includegraphics[width=0.6\textwidth]{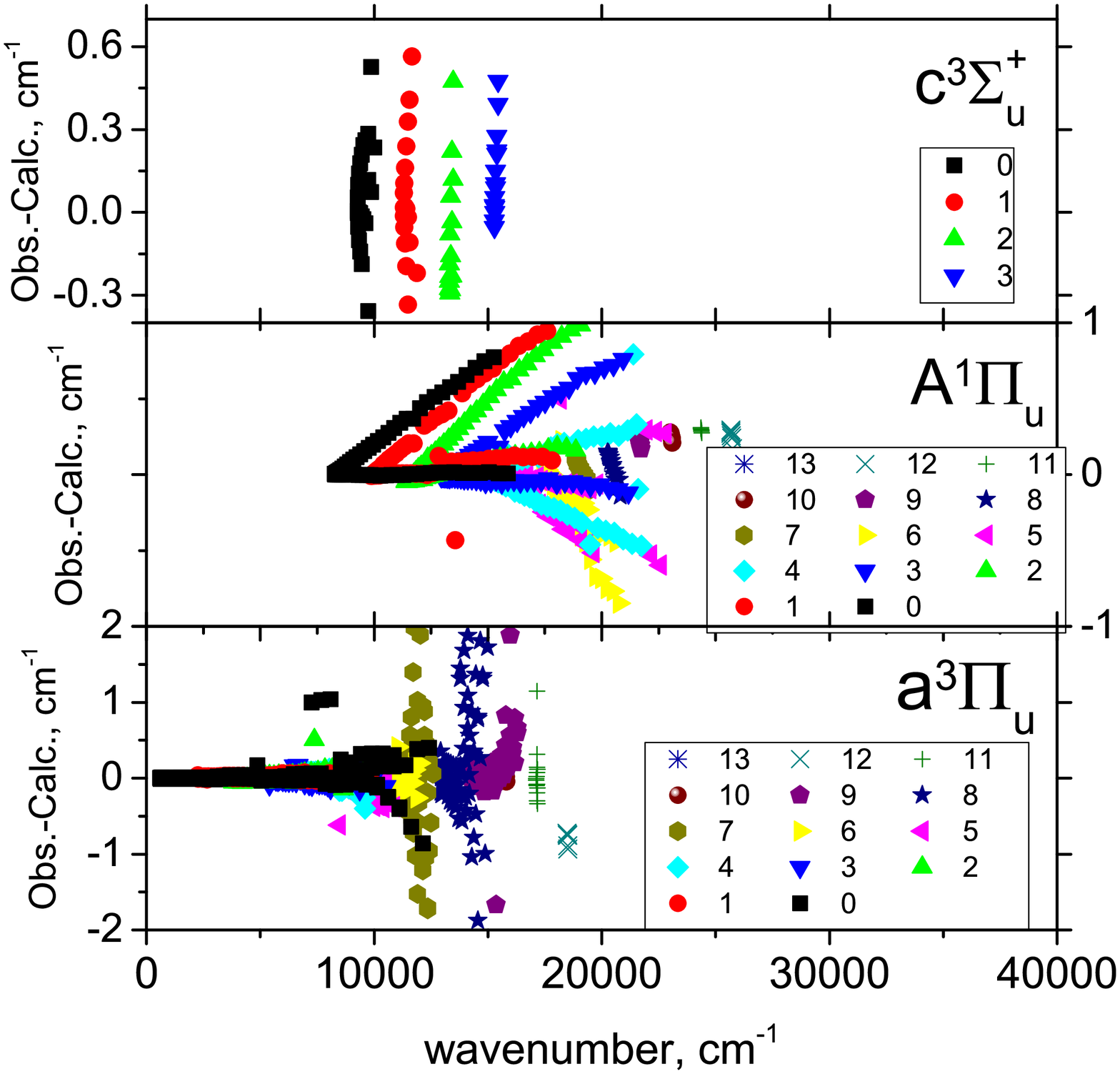}
\caption{Obs.$-$calc. residuals for individual vibronic bands of C$_2$: $u$-states.}
\label{f:obs-calc:u}
\end{figure}

The root-mean-square (rms) errors for individual vibronic states are listed in
Table~\ref{t:rsm}. An rms error as an averaged quantity does not fully reflect
the full diversity of the quality of the results, caused mostly by the
complexity of the couplings, which we could not fully describe.
Figs.~\ref{f:obs-calc} and \ref{f:obs-calc:u} present a detailed overview of the
Obs.$-$Calc. residuals for individual rovibronic states. Considering the avoided
crossings and other complexity of the system,
the generally small residues obtained represent a  huge achievement.
The final C$_2$ model is provided as \Duo\ input files as part of the
supplementary material  and can be also found at
\href{www.exomol.com}{www.exomol.com}.

\begin{table}
  \centering
  \caption{Rms errors (\cm) for obs.$-$calc. residuals achieved in the fit for individual vibronic states. }\label{t:rsm}
  \begin{tabular}{crccrccrccrccrccrccrccr}
\toprule
$v$&\multicolumn{1}{c}{$X$}&&
$v$&\multicolumn{1}{c}{$a$}&&
$v$&\multicolumn{1}{c}{$A$}&&
$v$&\multicolumn{1}{c}{$b$}&&
$v$&\multicolumn{1}{c}{$c$}&&
$v$&\multicolumn{1}{c}{$d$}&&
$v$&\multicolumn{1}{c}{$B$}&&
$v$&\multicolumn{1}{c}{$B^\prime$}\\
\hline
       0 &      0.01 &&     0 &      0.16 &&     0 &      0.26 &&     0 &      0.09 &&     0 &      0.19 &&     0 &      0.19 &&     0 &      0.01 &&     0 &      0.02\\
       1 &      0.01 &&     1 &      0.02 &&     1 &      0.33 &&     1 &      0.14 &&     1 &      0.74 &&     1 &      0.09 &&     1 &      0.07 &&     1 &      0.06\\
       2 &      0.16 &&     2 &      0.06 &&     2 &      0.35 &&     2 &      0.13 &&     2 &      0.23 &&     2 &      0.17 &&     2 &      0.06 &&     2 &      0.31\\
       3 &      0.33 &&     3 &      0.05 &&     3 &      0.26 &&     3 &      0.07 &&     3 &      0.21 &&     3 &      0.44 &&     3 &      0.05 &&     3 &      0.10\\
       4 &      0.36 &&     4 &      0.08 &&     4 &      0.22 &&     4 &      0.07 &&       &           &&     4 &      0.33 &&     4 &      0.03 &&       &          \\
       5 &      0.29 &&     5 &      0.13 &&     5 &      0.22 &&     5 &      0.14 &&       &           &&     5 &      0.28 &&     5 &      0.01 &&       &          \\
       6 &      0.31 &&     6 &      0.08 &&     6 &      0.33 &&     6 &      0.15 &&       &           &&     6 &      0.70 &&     6 &      0.04 &&       &          \\
       7 &      0.04 &&     7 &      0.72 &&     7 &      0.07 &&     7 &      0.10 &&       &           &&     7 &      0.31 &&     7 &      0.04 &&       &          \\
       8 &      0.01 &&     8 &      0.67 &&     8 &      0.10 &&     8 &      0.13 &&       &           &&     8 &      0.48 &&     8 &      0.18 &&       &          \\
       9 &      1.80 &&     9 &      0.32 &&     9 &      0.20 &&       &           &&       &           &&     9 &      1.00 &&       &           &&       &          \\
         &           &&    10 &      0.16 &&    10 &      0.25 &&       &           &&       &           &&    10 &      0.54 &&       &           &&       &          \\
         &           &&    11 &      0.34 &&    11 &      0.29 &&       &           &&       &           &&    11 &      0.82 &&       &           &&       &          \\
         &           &&    12 &      0.80 &&    12 &      0.26 &&       &           &&       &           &&    12 &      0.44 &&       &           &&       &          \\
         &           &&    13 &      2.43 &&    13 &      0.19 &&       &           &&       &           &&    13 &           &&       &           &&       &          \\
\bottomrule
\end{tabular}
\end{table}

\section{Line lists}

The line lists for three  isotopologues of the carbon dimer, $^{12}$C\2,
$^{13}$C\2\ and $^{12}$C$^{13}$C were computed using the refined model of the
eight lowest electronic states and the \ai\ transition DMCs. The line lists,
called 8states, cover the wavelength region up to 0.25~$\mu$m, $J=0\ldots 190$.
The upper state energy term values were truncated at 50~000~\cm. The lower state
energy threshold was set to 30~000~\cm\
so one can  assume that the other electronic states from the  region below
50~000~\cm\ (1{}$^{5}\Pi_g$, \Cstate, \Cpstate, \Dstate\ and \estate) are not
populated.
The vibrational excitation coverage for each electronic state was defined based
on the convergence and completeness to include all bound states below the first
dissociation limit.
We did not have problems with the numerical noise in production of overtone
intensities since they are simply forbidden, as are any transitions within the
same electronic states,
therefore no transition dipole moment cutoffs were applied.

The homonuclear molecule C$_2$ belongs to the infinite point symmetry group
\Dh{\infty}, which is also the group used in classification of the electronic
terms. The total rovibronic state spans a finite symmetry group \Dh{\infty}(M)
with four elements $E$ (the identity), $(12)$ (exchange of the identical nuclei),
$E^*$ (inversion), and $(12)^*$ \citep{97BuScHo,98BuJexx}. The
irreducible representations of \Dh{\infty}(M) are $\Sigma_{g}^+$,
$\Sigma_{g}^-$, $\Sigma_{u}^+$ and $\Sigma_{u}^-$.  For energy calculations
\duo\ uses  the  \Cv{\infty}(M) group to symmetrize its basis both for
homonuclear and heteronuclear systems. This group has two elements, $\Sigma^+$
and $\Sigma^-$, depending on whether the corresponding property is symmetric or
antisymmetric when the molecule is inverted. In case of homonuclear
$^{12}$C$_2$, the missing  symmetry is the permutation of the nuclei, which
introduces additional elements $g$ and $u$. This does not affect the energy
calculations as the absence of corresponding couplings between $g$ and $u$ is
guaranteed by construction.
However, it is important to use the proper symmetry for intensities mainly
due to the selection rules imposed by the nuclear spin statistics associated
with different irreducible representations.
For the homonuclear molecules like C$_2$ we therefore have to further classify
the rovibronic states according to $g$ and $u$. This is done  by simply adopting
the corresponding symmetry of the electronic terms.

The carbon atom $^{12}$C has a zero nuclear spin. This gives rise to
the zero statistical weights $g_{\rm ns}$ for the $\Sigma_{g}^-$ and
$\Sigma_{u}^+$ states, while the other two irreducible representations
have $g_{\rm ns}=1$. The statistical weights in case of $^{13}$C\2\
are $g_{\rm ns}=$ 1,1,3 and 3 for $\Sigma_{g}^+$, $\Sigma_{u}^-$, $\Sigma_{g}^-$ and
$\Sigma_{u}^+$, respectively.  For $^{13}$C$^{12}$C, all states have
 $g_{\rm ns}=2$. Note ExoMol follows the HITRAN convention \citep{jt692}
and includes the full nuclear-spin degenarcy in the partition function.
Other selection rules for the
electronic dipole transitions are:
$$
+ \leftrightarrow - , \quad g \leftrightarrow u, \quad J'-J'' = 0,\pm 1.
$$

The $^{12}$C\2\ line list contains 44~189 states and 6~080~920 transitions,
while the $^{13}$C\2\ and $^{12}$C$^{13}$C line lists comprise 94~003/13~361~992
and 91~067/1~2743~954 states/transitions, respectively.

The line lists include lifetimes and L\'ande-$g$ factors. Extracts from the line
lists are given in Tables~\ref{t:states} and \ref{t:trans}. In the final .states
file the theoretical (\Duo) energy term values were replaced with the
experimentally derived (MARVEL) values where available and indicated by a label
\verb!m!.

\begin{table}
\caption{Extract from the \texttt{.states} file for $^{12}$C$_2$.}
{\tt
\begin{tabular}{rrrrrrrrrrrrrc}
\toprule
$n$ & $\tilde{E}$ & $g_{\rm tot}$ & $J$ & $\tau$ & $g$-Land\'{e} & $+/-$ &  e/f & State & $v$ & $\Lambda$ &$\Sigma$ & $\Omega$ &m/d \\
\hline
     1  &      0.000000 &   1  &      0  &  Inf       &   0.000000  &    +   &  e   &    X1Sigmag+    &      0 &     0 &      0 &      0 &     m     \\
     2  &   1827.486182 &   1  &      0  &  1.71E+03  &   0.000000  &    +   &  e   &    X1Sigmag+    &      1 &     0 &      0 &      0 &     m     \\
     3  &   3626.681492 &   1  &      0  &  1.08E+03  &   0.000000  &    +   &  e   &    X1Sigmag+    &      2 &     0 &      0 &      0 &     m     \\
     4  &   5396.686466 &   1  &      0  &  1.34E+01  &   0.000000  &    +   &  e   &    X1Sigmag+    &      3 &     0 &      0 &      0 &     m     \\
     5  &   6250.149530 &   1  &      0  &  1.81E-05  &   0.000000  &    +   &  e   &    b3Sigmag-    &      0 &     0 &      0 &      0 &     m     \\
     6  &   7136.349911 &   1  &      0  &  6.52E-01  &   0.000000  &    +   &  e   &    X1Sigmag+    &      4 &     0 &      0 &      0 &     m     \\
     7  &   7698.252879 &   1  &      0  &  1.47E-05  &   0.000000  &    +   &  e   &    b3Sigmag-    &      1 &     0 &      0 &      0 &     m     \\
     8  &   8844.124324 &   1  &      0  &  1.94E-01  &   0.000000  &    +   &  e   &    X1Sigmag+    &      5 &     0 &      0 &      0 &     m     \\
     9  &   9124.177468 &   1  &      0  &  1.24E-05  &   0.000000  &    +   &  e   &    b3Sigmag-    &      2 &     0 &      0 &      0 &     m     \\
\ldots  &  \ldots       &      &         &  \ldots    &  \ldots     &        &      &                 &        &       &        &        &           \\
   309  &    619.642109 &   3  &      1  &  1.22E+04  &   0.892310  &    -   &  e   &      a3Piu      &      0 &    -1 &      0 &     -1 &     m     \\
   310  &    635.327453 &   3  &      1  &  4.27E+04  &  -0.392310  &    -   &  e   &      a3Piu      &      0 &    -1 &      1 &      0 &     m     \\
   311  &   2237.606490 &   3  &      1  &  3.54E+02  &   0.889334  &    -   &  e   &      a3Piu      &      1 &    -1 &      0 &     -1 &     m     \\
   312  &   2253.259339 &   3  &      1  &  1.41E+03  &  -0.389333  &    -   &  e   &      a3Piu      &      1 &    -1 &      1 &      0 &     m     \\
   313  &   3832.261242 &   3  &      1  &  1.38E+02  &   0.886238  &    -   &  e   &      a3Piu      &      2 &    -1 &      0 &     -1 &     m     \\
   314  &   3847.875681 &   3  &      1  &  5.44E+02  &  -0.386238  &    -   &  e   &      a3Piu      &      2 &    -1 &      1 &      0 &     m     \\
   315  &   5403.597872 &   3  &      1  &  7.81E+01  &   0.883039  &    -   &  e   &      a3Piu      &      3 &    -1 &      0 &     -1 &     m     \\
   316  &   5419.156714 &   3  &      1  &  3.06E+02  &  -0.383039  &    -   &  e   &      a3Piu      &      3 &    -1 &      1 &      0 &     m     \\
   317  &   6951.569985 &   3  &      1  &  2.65E+01  &   0.879743  &    -   &  e   &      a3Piu      &      4 &    -1 &      0 &     -1 &     m     \\
   318  &   6967.108683 &   3  &      1  &  4.06E+01  &  -0.379743  &    -   &  e   &      a3Piu      &      4 &    -1 &      1 &      0 &     m     \\
   319  &   8271.606854 &   3  &      1  &  1.31E-05  &   0.500002  &    -   &  e   &      A1Piu      &      0 &    -1 &      0 &     -1 &     m     \\
\bottomrule
\end{tabular}
}

\begin{tabular}{cll}
\\
             Column       &    Notation                 &      \\
\midrule
$i$:&   State counting number.     \\
$\tilde{E}$:& State energy in \cm. \\
$g$:&  Total statistical weight, equal to ${g_{\rm ns}(2J + 1)}$.     \\
$J$:& Total angular momentum.\\
$\tau$:& Lifetime (s$^{-1}$).\\
$g_J$:& Land\'{e} $g$-factor \\
$+/-$:&   Total parity. \\
$e/f$:&   Rotationless parity \citep{75BrHoHu.diatom,05Bernath.book}. \\
State:& Electronic state.\\
$\varv$:&   State vibrational quantum number. \\
$\Lambda$:&  Projection of the electronic angular momentum. \\
$\Sigma$:&   Projection of the electronic spin. \\
$\Omega$:&   $\Omega=\Lambda+\Sigma$, projection of the total angular momentum.\\
emp/calc:&   m=MARVEL, d=Duo. \\
\bottomrule
\end{tabular}
\label{t:states}

\end{table}

\begin{table}
\center
\tt
\caption{Extract of the first 13 lines from the $^{12}$C$_2$ \textsf{.trans} file.
Identification numbers \textit{f} and \textit{i} for
upper (final) and lower (initial) levels, respectively, Einstein-A coefficients denoted by \textit{A} (s$^{-1}$)
and transition frequencies $\nu$ (cm$^{-1}$). }
\begin{tabular}{rrrr} \hline\hline
\multicolumn{1}{c}{\textit{f}} & \multicolumn{1}{c}{\textit{i}} & \multicolumn{1}{c}{\textit{A}} & \multicolumn{1}{c}{$\nu$} \\
\hline
    2645  &      2025  &      3.2835E-10  &  140.623371    \\
    3199  &      3823  &      4.0106E-02  &  140.628688    \\
   10456  &     10728  &      8.7514E-07  &  140.643001    \\
    9518  &      9321  &      1.0017E-01  &  140.646479    \\
   12644  &     13248  &      2.8347E-11  &  140.659142    \\
   31380  &     31262  &      1.9673E-02  &  140.674836    \\
   19212  &     19072  &      7.0890E-08  &  140.695134    \\
   31818  &     31381  &      3.4496E-08  &  140.710566    \\
   13701  &     13087  &      2.6171E-09  &  140.710707    \\
    4772  &      4972  &      6.4432E-07  &  140.724342    \\
   24697  &     25214  &      9.6702E-08  &  140.724596    \\
    5111  &      5398  &      2.7821E-08  &  140.725422    \\
   14918  &     15183  &      6.6731E-07  &  140.728046    \\
\bottomrule
\end{tabular}
\label{t:trans}
\end{table}

\section{Results and discussion}

\subsection{Spectra}

All spectral simulations were performed using {\sc ExoCross} \citep{jt708}: our open-access Fortran~2003 code written to
work with molecular line lists.

\begin{figure}
\centering
\includegraphics[width=0.6\textwidth]{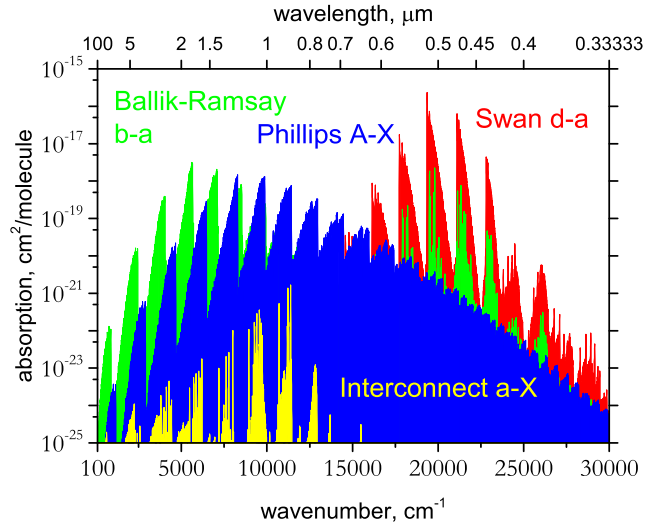}
\caption{Electronic bands of C$_2$ at $T=2000$~K using a Gaussian line profile with HWHM of 5~\cm. The forbidden interconnection band is shown using a light (yellow) colour. }
\label{f:bands:2000K}
\end{figure}

\begin{figure}
\centering
\includegraphics[width=0.6\textwidth]{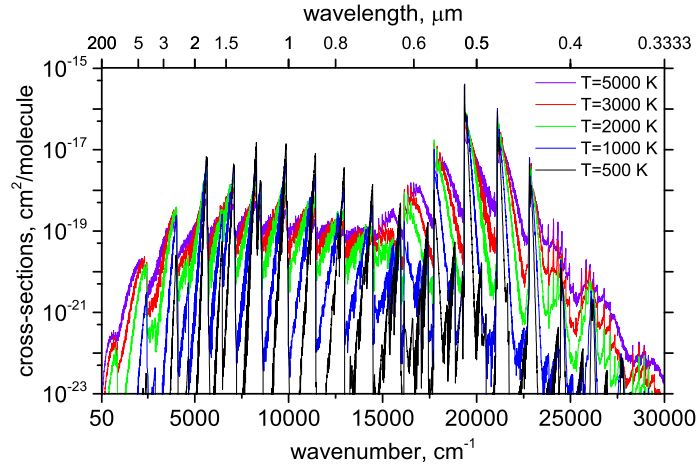}
\caption{Temperature dependence of the C\2\ cross sections using the 8states line list, from bottom to top: $T=$ 500, 1000, 2000, 3000 and 5000~K. }
\label{f:Temp}
\end{figure}

\begin{figure}
\centering
\includegraphics[width=0.6\textwidth]{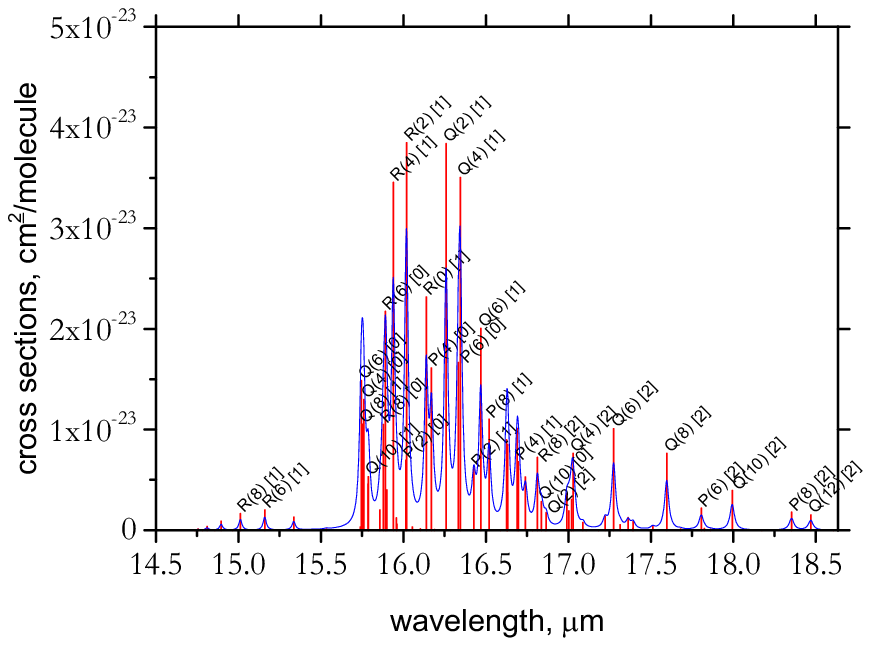}
\caption{\X\ -- \astate\ intercombination band of  C\2 in absorption at $T=100$~K. Individual 8states lines and
cross sections obtained using Lorentzian line profile of HWHM = 0.5~\cm are shown. The
major features are assigned; numbers in square brackets represent the $\Omega$ spin components $^3\Pi_{\Omega}$ of the \astate\ state.  }
\label{f:a-X}
\end{figure}

\begin{figure}
\centering
\includegraphics[width=0.49\textwidth]{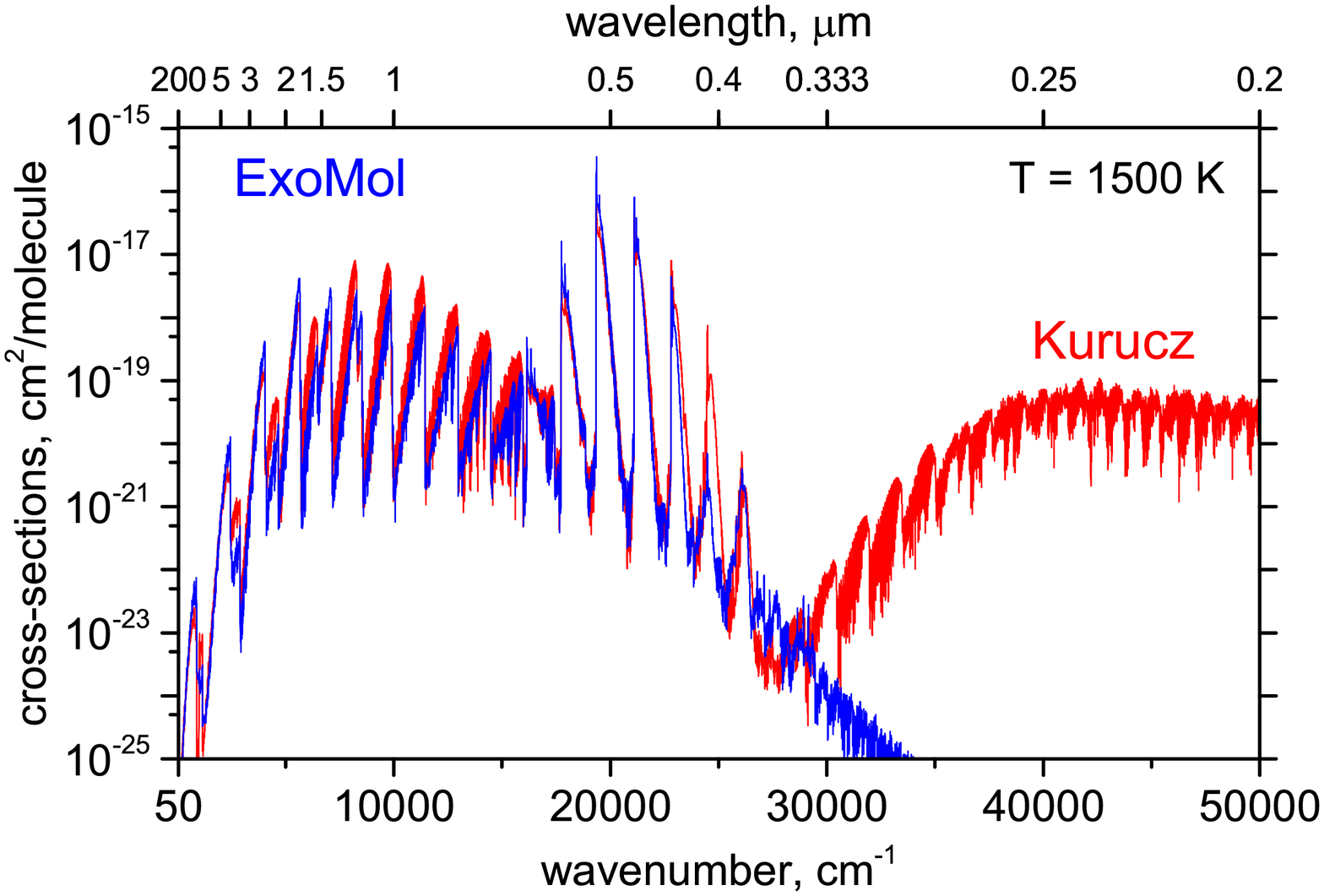}
\caption{8states comparing to Kurucz's line list  \protect\citep{11Kurucz.db} at $T=1100$~K. }
\label{f:Kurucz}
\end{figure}

Figure~\ref{f:bands:2000K} shows an overview of the electronic absorption
spectra of $^{12}$C\2\ at $T=2000$~K  and Fig.~\ref{f:Temp} shows the
temperature dependence of C\2\ absorption cross sections computed using the
8states line list.
The  singlet-triplet intercombination \X\ -- \astate\ band  is illustrated in
Figure~\ref{f:bands:2000K} as well as in Figure~\ref{f:a-X}.
Figure~\ref{f:Kurucz} compares the synthetic
absorption spectra of C\2\ at $T=1100$~K computed using our 8states line list with that by
\citet{11Kurucz.db}. The agreement is very good: \citet{11Kurucz.db}'s line list
has more extensive coverage, while ours is more accurate and complete below
40~000~\cm.

\begin{figure}
\centering
\includegraphics[width=0.49\textwidth]{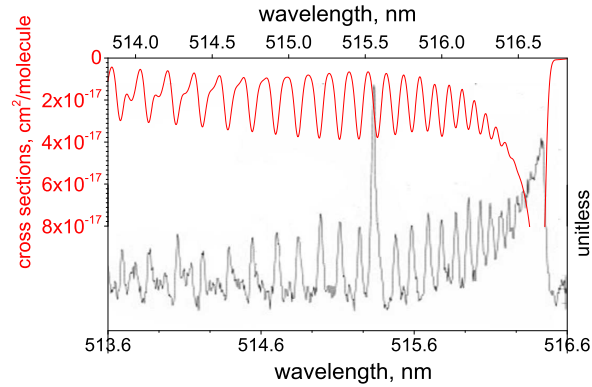}
\caption{The C\2\ Swan $(0,0)$ P branch band from the spectrum of V854 Cen on recorded on 1998 April 8 by \protect\citet{00RaLa} (lower trace),
compared to the theoretical spectrum at $T$=4625~K (quoted as rotation temperature by \protect\citet{00RaLa}) using a Gaussian line profile with HWHM=0.8~\cm\ (upper trace).
The star spectrum is red-shifted by 0.2195~nm.}
\label{f:V854}
\end{figure}

\begin{figure}
\centering
\includegraphics[width=0.49\textwidth]{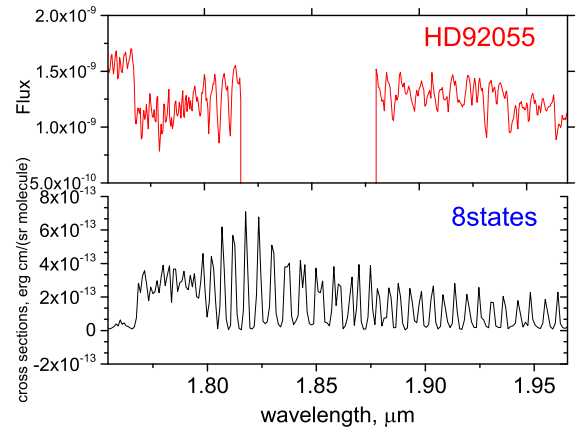}
\caption{Swan-band: Carbon-star HD~92055 spectra observed by \citet{09RaCuVa.dwarfs} at $R$=2000 compared to cross sections
simulated using $T=5000$~K  and a Gaussian profile with HWHM=1 \cm.}
\label{f:stars}
\end{figure}

\begin{figure}
\centering
\includegraphics[width=0.49\textwidth]{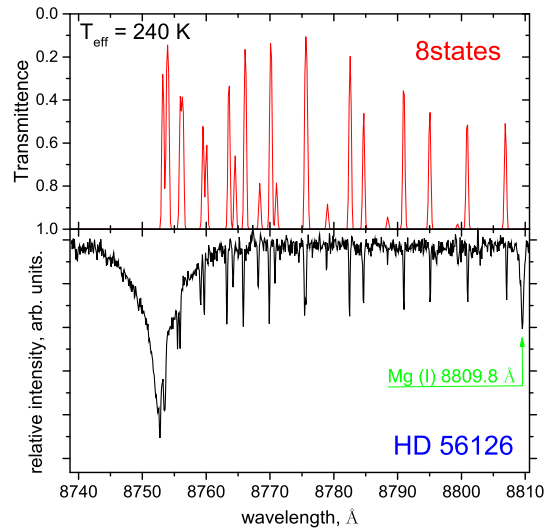}
\caption{The Phillips band  of C$_2$ at $T=240$~K using a Gaussian line profile with HWHM of 0.2~\cm\ (upper display) compared to the spectrum of the AGB remnants of HD~56126 observed by \protect\citet{96BaWaLa} (lower display). The HD~56126 spectrum was shifted to match the Mg(I) line to 8809.8~\AA. The theoretical transmittance spectrum is computed assuming the column amount of $10^{16}$~molecule/cm$^2$.
}
\label{f:AGB:Phillips}
\end{figure}

Figure~\ref{f:V854} shows a comparison of a Swan band-head (0,0) calculated
using our new line list and a stellar spectrum of V854 Cen \citep{00RaLa}. Figure~\ref{f:stars}
compares the theoretical flux spectrum of C\2\ with a stellar spectrum of the Carbon
star  HD~92055  \citep{09RaCuVa.dwarfs} at the resolving power $R$=2000.
Figure~\ref{f:AGB:Phillips} shows a simulated Philips band (2,0) compared to the
spectrum of AGB remnants of HD~56126 observed by \citet{96BaWaLa}. Similar
spectra of this band were reported by \citet{13ScZaPu} and \citet{12IsPaRe}.

\begin{figure}
\centering
\includegraphics[width=0.33\textwidth]{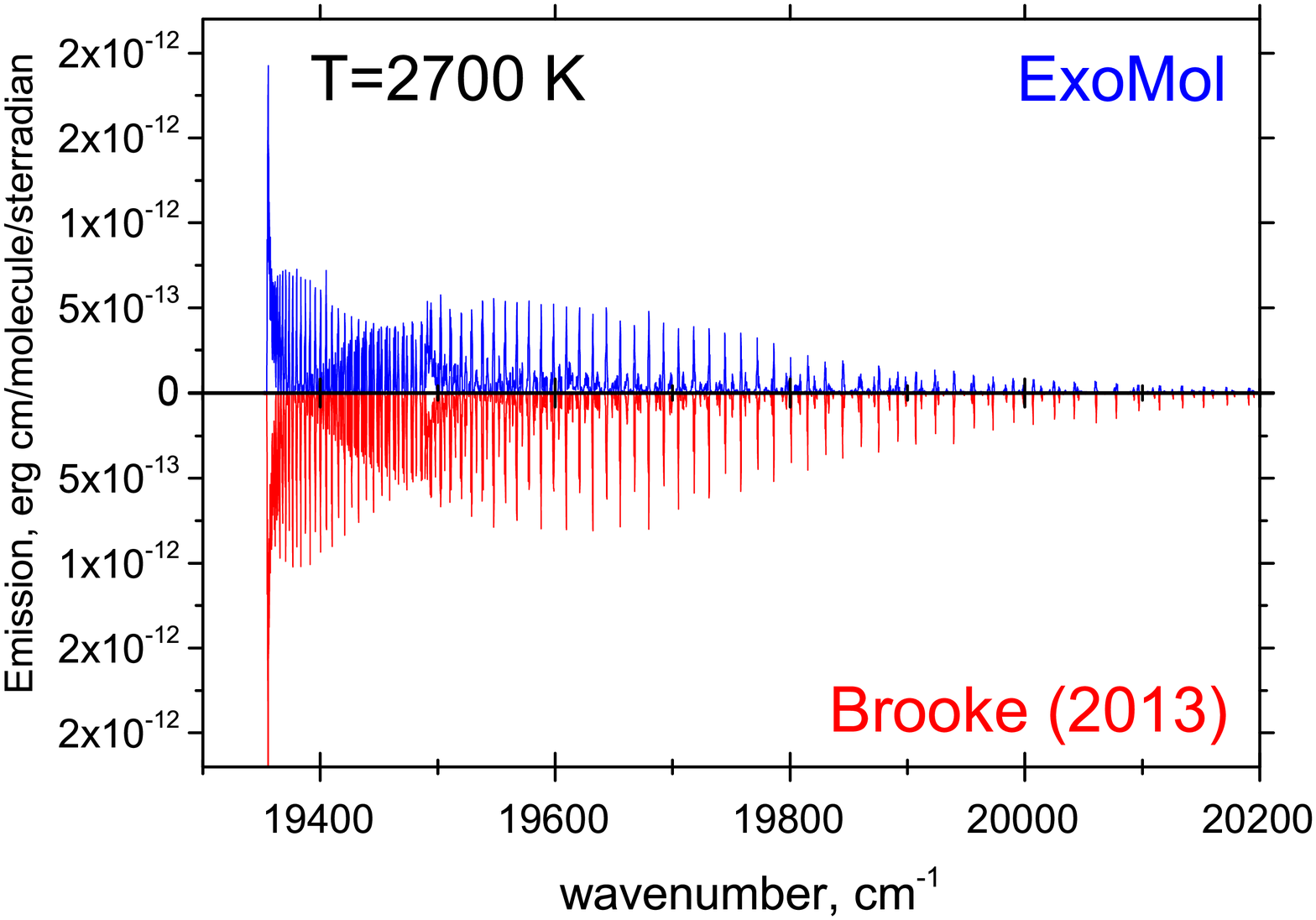}
\includegraphics[width=0.33\textwidth]{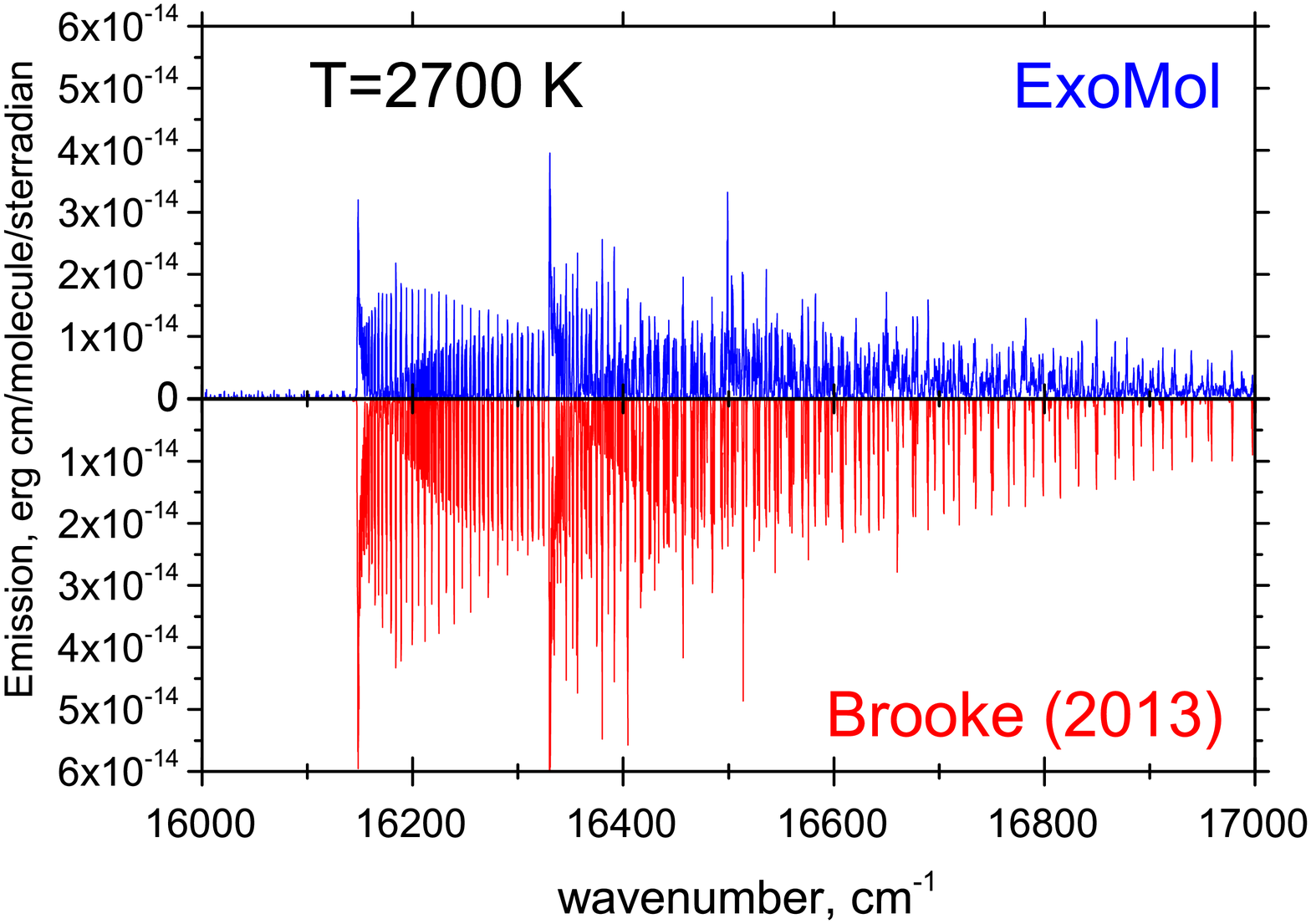}
\includegraphics[width=0.33\textwidth]{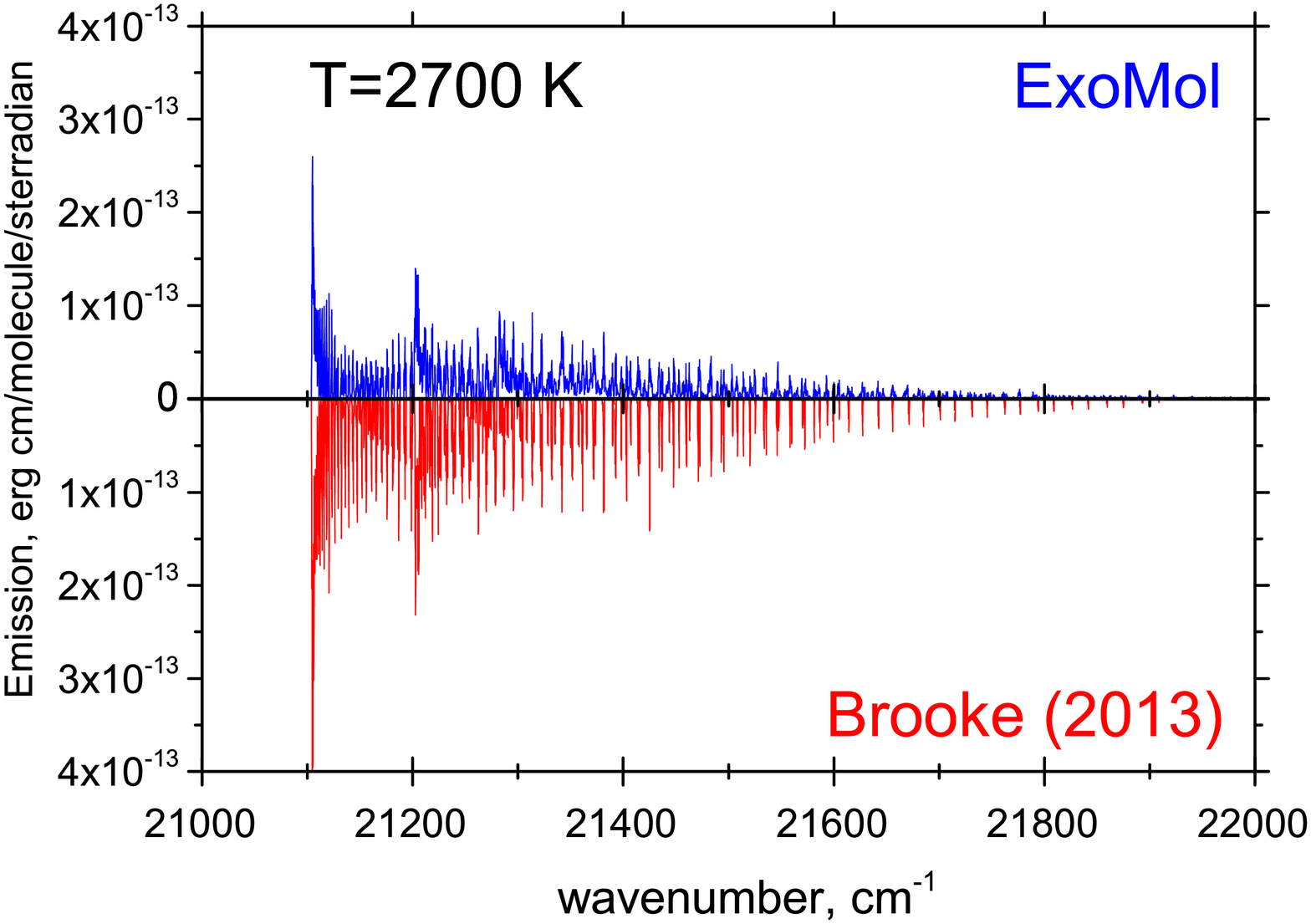}
\caption{The C\2\ Swan $(0,0)$, $(1,0)$ and $(0,1)$ bands in emission at $T=2700$~K simulated using  our line list and the empirical line list of \protect\citet{13BrBeScBa.C2} and a Gaussian line profile of HWHM=0.15~\cm.}
\label{f:Brooke:00:01:10}
\end{figure}

\begin{figure}
\centering
\includegraphics[width=0.49\textwidth]{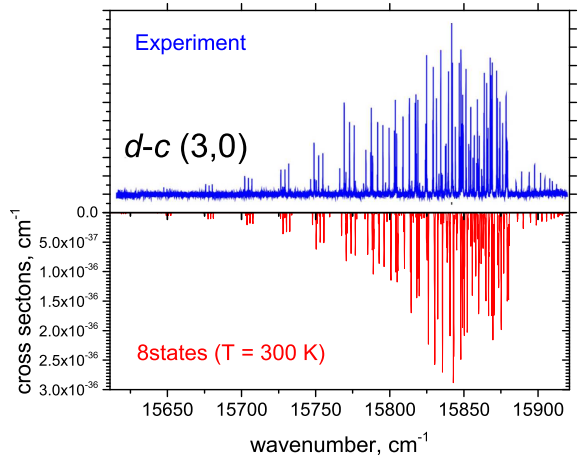}
\caption{The $d$--$c$ band  of C$_2$ at $T=300$~K using a Gaussian line profile with HWHM of 0.1~\cm\ (lower display) compared to the experimental spectrum of \citet{14NaEn.C2} (upper display). }
\label{f:d-c:300K}
\end{figure}

 Figure~\ref{f:Brooke:00:01:10} gives detailed, high resolution emission spectra of the (0,0), (1,0) and (0,1) Swan bands computed using our line list and the empirical line list by \citet{13BrBeScBa.C2}.
Figure~\ref{f:d-c:300K} shows a simulation of the $d$--$c$ (3,0) band of C\2\ compared to the experiment by \citet{14NaEn.C2}.

\begin{figure}
\centering
\includegraphics[width=0.75\textwidth]{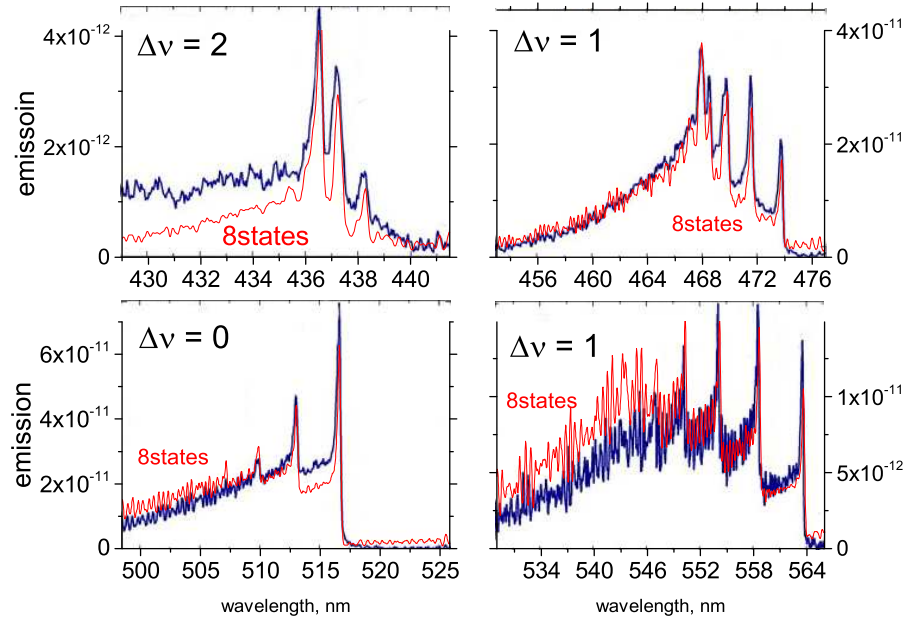}
\caption{Swan-bands: a plasma spectrum of C\2\ by \protect\cite{13AlHaHa} and our emission spectrum (erg cm sr$^{-1}$molecule$^{-1}$) at $T=8000$~K
for four different vibronic bands as specified in each panel.
A Gaussian profile with the HWHM=4~\cm\ was used. }
\label{f:13AlHaHa}
\end{figure}

Figure~\ref{f:13AlHaHa} shows a plasma spectrum of C\2\ recorded by  \citet{13AlHaHa} compared to 8states emission cross sections at $T=$8000~K.

\subsection{Isotopic shifts}

\begin{figure}
\centering
\includegraphics[width=0.55\textwidth]{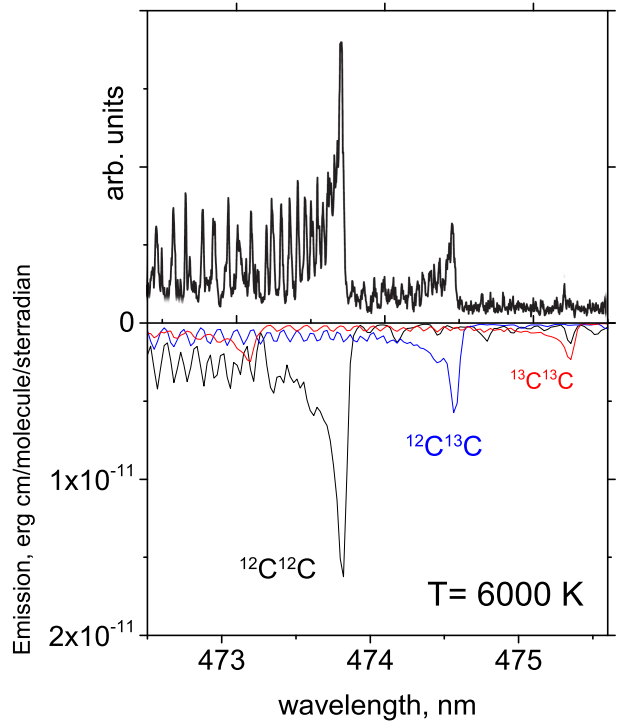}
\caption{Isotopic shift in the spectra of C\2, Swan (1,0) for three isotopologues, $^{12}$C\2, $^{12}$C$^{13}$C and $^{13}$C\2. Upper display: The experimental spectrum of a laser-induced plasma by \protect\citet{14DoChXi}. Lower display: The theoretical emission spectrum at $T=$ 6000~K computed using the Gaussian profile with the HWHM=1~\cm\ (blue-shifted by 0.1~nm to match the band heads). The theoretical abundances of $^{12}$C$^{13}$C and $^{13}$C\2 were scaled to match the experimental intensities by the factors 0.34 and 0.14, respectively. } \label{f:14DoChXi}
\end{figure}

Figure~\ref{f:14DoChXi} shows the effect of the isotopic substitution on the
vibronic spectra of C\2\ for the (1,0)  Swan  band of
$^{12}$C\2, $^{12}$C$^{13}$C and $^{13}$C\2\ at $T=6000$~K  compared to the experimental, laser-induced plasma spectrum of \citet{14DoChXi}.

\subsection{Partition function}

\begin{figure}
\centering
\includegraphics[width=300pt]{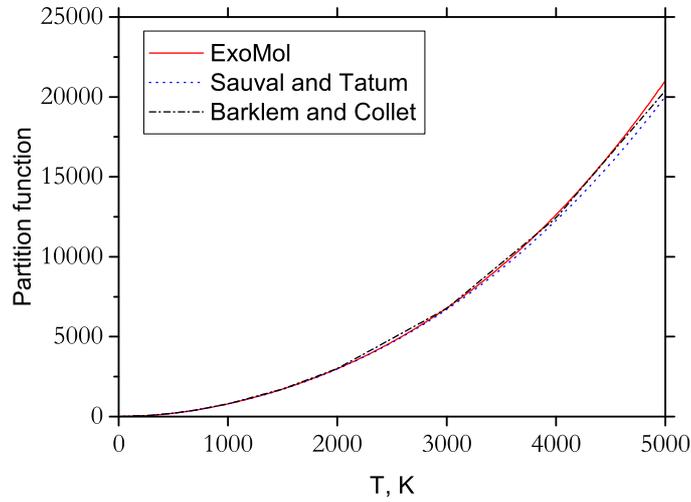}
\caption{Temperature dependence of the partition functions of C\2\ computed using our line list and compared to that  by \protect\citet{84SaTaxx.partfunc} and \protect\citet{16BaCoxx.partfunc}.}
\label{f:pf}
\end{figure}

As part of the line list package and as supplementary material we also report partition functions of the three C\2\ isotopologues up to 10~000~K at 1 K intervals. Figure~\ref{f:pf} shows the partition functions of $^{12}$C\2\ computed using the 8states line list and compared to that by \citet{84SaTaxx.partfunc} and \citet{16BaCoxx.partfunc}. All three partition functions are in a good agreement.

We have also fitted the partition functions to the function form of \citet{jt263}:
\begin{equation}\label{eq:fit1d}
\log_{10}Q(T) = \sum\limits^{9}_{n=0} a_n (\log_{10}T)^n.
\end{equation}
Table~\ref{t:pf} gives the expansion coefficients for all three isotopologues considered, which reproduce the our partition functions within 1~\%\ (relative values) for $T>300$~K and within $\sim$1 (absolute values) for $T<$ 300~K.

\begin{table}
\caption{Fitting parameters used to represent the partition functions of C\2,
see Eq.~(\ref{eq:fit1d}). Fits are valid for temperatures up to 10000~K.
}
\label{t:pf}
\begin{center}
\begin{tabular}{lrrr}
\hline
\hline
Parameter & \multicolumn{1}{c}{$^{12}$C$^{12}$C}          &
\multicolumn{1}{c}{$^{13}$C$^{13}$C}           &
\multicolumn{1}{c}{$^{12}$C$^{13}$C}               \\
\hline
$       a_0     $&$       7.6691867335   $&$         -0.0426980211   $&$          0.2401679500       $\\
$       a_1     $&$     -45.6411466388   $&$         16.5774126902   $&$         15.5801797784       $\\
$       a_2     $&$      91.4980228540   $&$        -66.5674043182   $&$        -65.2111429415       $\\
$       a_3     $&$     -88.7302064590   $&$        113.4659696600   $&$        112.5717888350       $\\
$       a_4     $&$      48.2113438335   $&$       -102.5476534740   $&$       -102.3357156140       $\\
$       a_5     $&$     -15.3601370784   $&$         54.3985236316   $&$         54.4790813354       $\\
$       a_6     $&$       2.8436843218   $&$        -17.5150320571   $&$        -17.5862991363       $\\
$       a_7     $&$      -0.2820217958   $&$          3.3710897879   $&$          3.3919572101       $\\
$       a_8     $&$       0.0115099797   $&$         -0.3570362951   $&$         -0.3599156561       $\\
$       a_9     $&$                      $&$          0.0160186260   $&$          0.0161757189       $\\
\hline
\end{tabular}
\end{center}
\end{table}

\subsection{Lifetime}

\begin{figure}
\centering
\includegraphics[width=0.6\textwidth]{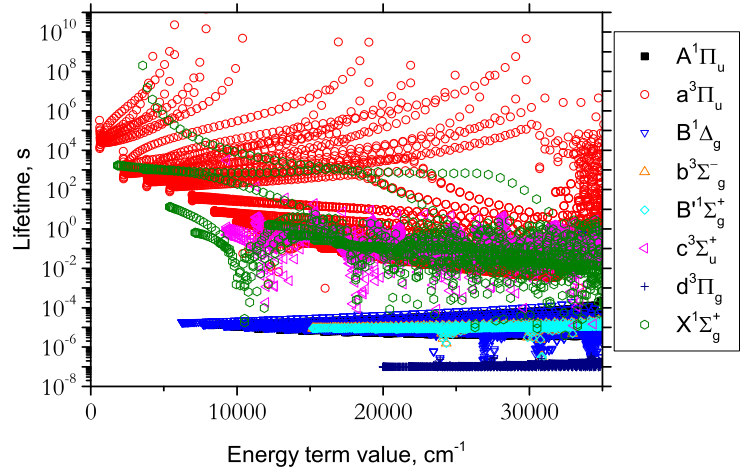}
\caption{Lifetimes of rovibronic states of C\2. The black solid squares representing the \A\ state values are hidden behind the blue open triangles of the \bstate\ state.}
\label{f:lf}
\end{figure}

We have computed life times of C\2\ for all rovibronic states below
30~000~\cm. These are compared to the experimental and theoretical
values by
\citet{69Smithx.C2,75CoNich.C2,76CuEnEr.C2,85BaBeHu,86BaBeBiMe,88NaCoDo.C2}
and \citet{95ErIwam.C2}. The agreement is good and comparable to the previous
\ai\ values (Davidson corrected MRCI/aug-ccpV6Z level) obtained by
\citet{07ScBaxx.C2} and \citet{07KoBaSc.C2}. The lifetimes are also
illustrated in Fig.~\ref{f:lf}. The rather unusual long life times of
the lower rovibronic states of \astate\ are explained by crossing with
the lower states of \X\ at about $J=50$, where the \astate\ rovibronic
states are lower than the \X\ rovibronic states. Up to $J=48$ the
lowest state in each $J$-manifold is \X, $v=0$, which has an infinite
lifetime. Starting from $J=50$ the lowest rovibronic state with the
infinite lifetime is \astate, $v=0$, $|\Omega|=1$.  By $J=125$ there
are six infinitively living \astate\ rovibronic states ($v=0,1$).

\begin{table}
\caption{Experimental and calculated lifetimes of C$_2$ vibronic states: note the different time scales used to represent the lifetimes associated with
different electronic states. Previous calculations have only considered limited decay routes for each state; these are noted in the
footnote. The results suggest that these assumptions were justified.}
\label{tab:LF}
\footnotesize
\begin{tabular}{llrrrrrrl}
\hline\hline
State (units)/$v$     &   &  0     &           1       &        2  &                 3          &       4            &   5       &   Source                   \\
     \hline
     \\
\A      & Calc.$^b$         &$          13 $&$         10.4 $&$        8.77 $&$                7.65 $&$          6.84 $&$        6.22 $& \citet{07ScBaxx.C2}                   \\
(ms)           & Calc.$^b$         &$        16.6 $&$         13.1 $&$        11.0 $&$                9.55 $&$          8.50 $&$        7.71 $& \citet{07KoBaSc.C2}                   \\
               & Exp.         &$             $&$              $&$             $&$         6.8\pm 2.0  $&$   7.1\pm 1.1  $&$             $& \citet{95ErIwam.C2}                   \\
               & Exp.         &$ 13.4\pm 2.5 $&$ 15.4\pm 4.0  $&$ 14.4\pm 2.0 $&$        12.0\pm 2.0  $&$  10.7\pm 2.0  $&$ 7.9\pm 2.0  $& \citet{85BaBeHu}                      \\
               & Exp.             &$  18.5\pm 3  $&$              $&$             $&$          11.4\pm 2  $&$               $&$             $& \citet{86BaBeBiMe}                    \\
          &  Calc.        &$       13.13 $&$        10.62 $&$        8.95 $&$                7.77 $&$          6.93 $&$      6.911 $&           This work                            \\
\\
\dstate      & Calc.$^c$        &$        95.1 $&$         96.7 $&$        99.1 $&$                 102 $&$           107 $&$         113 $& \citet{07ScBaxx.C2}                   \\
(ns)             & Calc.$^c$        &$          98 $&$         99.8 $&$       102.4 $&$                 106 $&$         110.9 $&$       118.2 $& \citet{13BrBeScBa.C2}                 \\
               & Calc.$^c$        &$      103.20 $&$       104.97 $&$      107.86 $&$              111.66 $&$        116.77 $&$      123.23 $& \citet{07KoBaSc.C2}                   \\
               & Exp.         &$ 101.8\pm 4. $&$ 96.7\pm 5.2  $&$  104\pm 17  $&$                     $&$               $&$             $& \citet{88NaCoDo.C2}                   \\
               & Exp.         &$  106\pm 15  $&$   105\pm 15  $&$             $&$                     $&$               $&$             $& \citet{86BaBeBiMe}                    \\
               & Exp.         &$   123\pm 6  $&$    124\pm 6  $&$   130\pm 6  $&$           128\pm 6  $&$     131\pm 6  $&$  137\pm 10  $& \citet{76CuEnEr.C2}                      \\
\          & Calc.        &$       99.37 $&$       101.62 $&$      104.71 $&$              108.69 $&$        115.18 $&$       119.3 $& This work                                       \\
\\
\bstate      & Calc.$^d$        &$       17.83 $&$        14.59 $&$       12.47 $&$               11.09 $&$          9.88 $&$        9.38 $& \citet{07ScBaxx.C2}                   \\
(ms)           & Exp.$^e$       &$    17.2         $&$              $&$         $&$       $&$               $&$             $& \citet{75CoNich.C2} \\
               & Calc.        &$       17.00 $&$        14.46 $&$       12.68 $&$               11.46 $&$         10.39 $&$      9.612 $&     This work                                   \\
\hline \hline
\end{tabular}

\begin{tabular}{l}
$^a$ \Dstate\ $\rightarrow$ \X, \Bp\ and \Cstate\ only considered.\\
$^b$ \A\ $\rightarrow$ \X\ only considered.\\
$^c$  \dstate\ $\rightarrow$ \astate\ only considered.\\
$^d$  \bstate\ $\rightarrow$ \astate\ only considered.\\
$^e$ Average value for a range of vibrational states
\end{tabular}
\end{table}

\section{Conclusions}

New empirical rovibronic line lists for three isotopologues of C\2\
($^{12}$C$_2$,$^{13}$C$_2$ and $^{12}$C$^{13}$C) are presented. These
line lists, called 8states, are based on high level \ai\ (MRCI)
calculations and empirical refinement to the experimentally derived
energies of $^{12}$C\2.  The line lists cover eight lowest electronic
(singlet and triplet) states \allstates\ fully coupled in the nuclear
motion calculations through spin-orbit and electronic angular momentum
curves and complemented by empirical curves representing different
corrections (Born-Oppenheimer-breakdown, $\Lambda$-doubling, spin-spin
and spin-rotation). The line lists should be complete up to about
30~000~\cm\ with the energies stretching up to 50~000~\cm. In order to
improve the accuracy of the line positions, where available the
empirical energies were replaced by experimentally derived MARVEL
values. The line lists were benchmarked against high temperature
stellar and plasma spectra. Experimental lifetimes were especially
important for assessing our absolute intensities as well as the
quality of the underlined \ai\ dipole moments of C\2\ used.  The line
lists, the spectroscopic models and the partition functions are
available from the CDS
(\href{http://cdsarc.u-strasbg.fr}{http://cdsarc.u-strasbg.fr}) and
ExoMol (\href{www.exomol.com}{www.exomol.com}) databases.

\section{Acknowledgements}

We thank Andrey Stolyarov, Timothy W. Schmidt  and George B. Bacskay for help at different stages of the project. This work was supported by the UK Science and Technology Research Council (STFC) No. ST/M001334/1 and the COST action MOLIM No. CM1405.  This work made extensive use of UCL's Legion  high performance computing facility. A part of the calculations were performed using DARWIN, high performance computing facilities provided by DiRAC for particle physics, astrophysics and cosmology and supported by  BIS National E-infrastructure capital grant ST/J005673/1 and STFC grants ST/H008586/1, ST/K00333X/1.

\label{lastpage}

\clearpage
\bibliographystyle{mnras}

\end{document}